\documentclass[twocolumn, showkeys, showpacs, prb, floatfix]{revtex4}
\usepackage{amsfonts, amsmath}
\usepackage[dvips]{graphicx}
\usepackage{url}

\newtheorem{1}{Theorem}
\newtheorem{2}[1]{Theorem}
\newtheorem{3}[1]{Theorem}
\newtheorem{4}[1]{Theorem}
\newtheorem{Pr1}{Proposition}

\newtheorem{5}[1]{Theorem}
\newtheorem{Cor1}{Corollary}
\newtheorem{6}[1]{Theorem}

\begin{document}

\title{Existence of short-time approximations of any polynomial order for the computation of density matrices by path integral methods\footnote{This article, except for Appendix~B, is an improved version of Chapter~IV of the author's Ph.D. dissertation, which has been submitted in partial fulfillment of the requirements for the Degree of Doctor of Philosophy in the Department of Chemistry at Brown University.}}

\author{Cristian Predescu}
 
\affiliation{
Department of Chemistry, Brown University, Providence, Rhode Island 02912
}
\date{\today}
\begin{abstract}
In this article, I provide significant  mathematical evidence in support of the existence of direct short-time approximations of any polynomial order for the computation of density matrices of physical systems described by arbitrarily smooth and bounded from below potentials. While for  Theorem~\ref{th:2}, which is ``experimental,'' I only provide a ``physicist's'' proof, I believe the present development is mathematically sound. As a verification, I explicitly construct two short-time approximations to the density matrix having convergence orders $3$ and $4$, respectively. Furthermore, in Appendix~B, I derive the convergence constant for the trapezoidal Trotter path integral technique. The convergence orders and constants are then verified by numerical simulations. While the two short-time approximations constructed are of sure interest to physicists and chemists involved in Monte Carlo path integral simulations, the present article is also aimed at the mathematical community, who might find the results interesting and worth exploring.  I conclude the paper by discussing the implications of the present findings with respect to the solvability of the dynamical sign problem appearing in real-time Feynman path integral simulations. 
\end{abstract}

\maketitle
\section{Introduction}
In the path integral formulation, the density matrix of a thermodynamic system is expressed as  the expected value of a functional of the Brownian motion by means of the Feynman-Kac formula\cite{Fey48, Kac51, Sim79}  
\begin{equation}
\label{eq:I1}
\frac{\rho(x,x';\beta)}{\rho_{fp}(x,x';\beta)}=\mathbb{E}\exp\left\{-\beta\int_{0}^{1}\! \!  V\Big[x_r(u)+\sigma B_u^0 \Big] d u\right\}.
\end{equation}
Here, $\rho(x,x';\beta)$ is the density matrix for a one-dimensional canonical system characterized by the inverse temperature $\beta=1/(k_B T)$ and made up of identical particles of mass $m_0$  moving in the potential $V(x)$. 
The stochastic element that appears in Eq.~(\ref{eq:I1}), $\{B_u^0,\, 0 \leq u \leq 1\}$, is a so-called standard Brownian bridge, defined as follows: if  $\{B_u,\, u\geq
0\}$ is a standard Brownian motion starting at zero, then the Brownian
bridge is the stochastic process~$\{B_u, 0 \leq u \leq 1 |\,B_1=0\}$,
i.e., a Brownian motion conditioned on the event~$B_1=0$. A Brownian bridge can be realized as  the process $\{B_u - u B_1, 0 \leq u \leq 1\}$.\cite{Dur96} For additional information on Brownian motion and its relation to the Feynman-Kac formula, the reader is advised to consult Appendix~A as well as the cited bibliography. To complete the description of Eq~(\ref{eq:I1}), we set $x_r(u)=x+(x'-x)u$ (called the reference path), $\sigma= (\hbar^2\beta  /m_0)^{1/2}$, and let $\rho_{fp}(x,x';\beta)$ denote the density matrix for a similar free particle.

 The $d$-dimensional generalization of the Feynman-Kac formula is rather trivial. One just considers an independent Brownian bridge for each additional degree of freedom. To keep the notation simple, in this article we shall work exclusively with one-dimensional systems. However, the reader should notice that the main results of the paper remain true or have straightforward generalizations for systems of arbitrary dimensionality.

In actual simulations, the Feynman-Kac formula is almost always used in conjunction with Monte Carlo  integration methods\cite{Cep95} and, for this purpose, one needs to construct rapidly convergent finite-dimensional approximations to the stochastic integral described by Eq.~(\ref{eq:I1}). Ideally, such approximations should require knowledge of the potential only for the computation of the density matrix or the partition function of the physical system. This type of methods will be called direct methods. The main question we address in the present article concerns the rate of convergence of a class of discretization techniques as measured against the number of variables utilized for path parameterization. Throughout the paper, we assume that the potential $V(x)$ is an infinitely differentiable and bounded from below function. 

Until recently, the fastest  direct method available (as order of convergence) has been the trapezoidal Trotter discrete path integral (DPI) method.\cite{Tro59,Rae83} The technique is usually derived by means of the Lie-Trotter product formula and an appropriate short-time high-temperature approximation. The formal asymptotic convergence of the trapezoidal Trotter DPI method and of  related DPI techniques was extensively studied by Suzuki\cite{Suz91, Suz85} and was found to be $O(1/n^2)$. I shall comment more on this method in Section~II.A. With the introduction of the random series implementation of the Feynman-Kac formula,\cite{Pre02} faster methods became available. More precisely, two examples of direct path integral techniques constructed around the L\'evy-Ciesielski and the Wiener-Fourier random series representations of the Brownian motion and  pertaining to the general class of reweighted random series techniques were shown to have $O(1/n^3)$ asymptotic convergence.\cite{Pre03, Pre03b} In a recent Monte Carlo simulation,\cite{Pre03c}  the superior convergence of the reweighted methods proved to be crucial for the accurate determination of the potential, kinetic, and total energies of a highly quantum mechanical Lennard-Jones cluster made up of $22$ molecules of hydrogen at a temperature of $6\,\text{K}$.

In this article, I try to argue that in fact, for infinitely differentiable potentials $V(x)$, there might exist direct short-time high-temperature approximations of arbitrary polynomial convergence order.  The construction of such approximations is based upon an ``experimental'' theorem on the pointwise convergence of Lie-Trotter product formulas. This theorem is presented in Section II.A, where it is used to derive a set of  functional equations that  short-time approximations must satisfy in order to have a given convergence order. Unlike standard approaches based upon the construction of ``effective'' potentials,\cite{Fey65, Gia85,  Fey86, Tak84, Cep95, Dol85} the short-time approximations we consider in the present article  are based on carefully designed finite-dimensional approximations to the Brownian motion entering the Feynman-Kac formula. The potential itself is left unchanged. It is for this reason that the set of equations mentioned above do not depend upon the potential. The equations can be solved once for a given order and their (not unique) solutions can be tabled and used in actual computations for all potentials.

The main mathematical problem that is left unsolved in this article is the existence of finite-dimensional approximations to the Brownian motion that satisfy the functional equations for a given convergence order. To support the idea that such solutions exist, I explicitly construct two short-time approximations to the density matrix having convergence orders $3$ and $4$, respectively. A solution for the order $3$ has been previously derived,\cite{Pre03} but the one I construct in the present paper utilizes fewer path variables and fewer quadrature points.  The solution for the order $4$ is derived as evidence that the general problem of constructing finite-dimensional approximations of arbitrary order is positively solvable. The fourth order method has numerical requirements similar to the trapezoidal Trotter method (as ratio number of calls to the potential over number of path variables). The method has been recently utilized in the study of the heat capacity of the $\text{Ne}_{13}$ cluster.\cite{Pre03f} 

In Section~V, I verify by numerical simulations the asymptotic convergence of the two short-time approximations discussed above. The definite agreement with the theoretical predictions is interpreted as proof that the theoretical development in the present article is mathematically sound. I conclude the paper by speculating that sequences of short-time approximations for increasing convergence orders (if they exist) may provide exponentially fast approximations for  imaginary-time ``propagated'' wavefunctions, as measured against the number of path variables. I then analyze the implications of this hypothesis with respect to the solvability of the dynamical sign problem  for  real-time Feynman path integrals on a classical computer.

In Appendix~B, I derive the convergence constant for the celebrated trapezoidal Trotter path integral technique. The convergence constant is verified by numerical simulations. The excellent agreement between theory and simulation is interpreted as further evidence that  Theorem~2 is a valid mathematical statement (perhaps after further restrictions on its hypothesis).

 \section{Product approximations}
In the first part of this section, I review the classical results of Suzuki concerning the order of convergence of a special family of short-time approximations. These results serve illustrate the main difficulties regarding the construction of short-time approximations having convergence orders higher than $2$. I then state a theorem  concerning the pointwise convergence of Lie-Trotter product formulas and discuss its implications with respect to the design of short-time approximations having superior convergence orders.   In Section~II.B, I introduce a special class of short-time approximations constructed by replacing the Brownian motion appearing in the Feynman-Kac formula with appropriate finite-dimensional Gaussian processes. The functions utilized in the construction of these finite-dimensional Gaussian processes will become the unknown variables for the systems of functional equations controlling the orders of convergence of the associated short-time approximations. These systems of functional equations are derived in Section~III.

\subsection{A convergence theorem for product formulas}

One of the most fruitful approaches to constructing finite-dimensional approximations to the quantum mechanical density matrix was given by Trotter.\cite{Tro59} It exploits the fact that $\{e^{-\beta H}; \beta > 0\}$ is a semigroup of operators on $L^2(\mathbb{R})$, so that 
\begin{equation}
\label{eq:II1}
e^{-(\beta_1 +\beta_2)H}= e^{-\beta_1 H}e^{-\beta_2 H}
\end{equation} 
or, in coordinate representation,
\begin{equation}
\label{eq:II2}
\langle x|e^{-(\beta_1 +\beta_2)H}| x' \rangle= \int_{\mathbb{R}} d z \langle x| e^{-\beta_1 H}|z\rangle \langle z|e^{-\beta_2 H}|x'\rangle.
\end{equation} 
(In this work, the Hamiltonian, the kinetic operator, and the potential operator are denoted by the symbols $H$, $K$, and $V$, respectively.)  The Trotter approximation theorem states that 
\[
e^{-\beta H} = \lim_{n \to \infty} \left[e^{-\beta K / n} e^{-\beta V / n}\right]^n
\]
in the sense of strong operator convergence. The quantity
\[
e^{-\beta K / n} e^{-\beta V / n}
\]
is called a short-time high-temperature approximation of the exact density matrix operator $e^{-\beta H/n}$.  

There has been a lot of research on the rate of convergence of the above approximation or of similar Trotter-like formulas. Of particular significance is Suzuki's work,\cite{Suz91} which treats the more general problem based on  short-time approximations of the form
\begin{eqnarray}
\label{eq:II3}
e^{-\beta (K+V)}\nonumber= &&e^{-a_0 \beta V}e^{-b_1 \beta K}e^{-a_1 \beta V}\ldots \\ &&\ldots e^{-b_l \beta K} e^{-a_{l} \beta V} [1+{O}(\beta^{\nu+1})], 
\end{eqnarray} 
where the sequences of non-negative real numbers $a_0, a_1, \ldots, a_l$ and $b_1, b_2, \ldots, b_l$ are palindromic and sum to $1$. Following Suzuki, a short-time approximation $f_\nu(K,V;\beta)$ is called of order $\nu$ if
\[e^{-\beta(K+V)}=f_\nu(K,V;\beta)[1+{O}(\beta^{\nu+1})].\]
In this case,\cite{Suz85} 
\begin{equation}
\label{eq:II4}
e^{-\beta(K+V)}=\left[f_\nu\left(K,V;\frac{\beta}{n}\right)\right]^n \left[1+{O}\left(\frac{\beta^{\nu+1}}{n^\nu}\right)\right].
\end{equation}
[To be rigorous, I mention that Eq.~(\ref{eq:II4}) has been proved for bounded operators $A$ and $B$. The respective theorem states that the operator norm error of the final $n$-term Lie-Trotter product formula decays as fast as $1/n^\nu$. However, experience shows that the orders of convergence are correctly predicted even for the unbounded operators $K$ and $V$. Moreover, if the non-existence theorem discussed below is true for bounded operators, then it  is also true for the more general class of unbounded operators.]

The more general splitting formula given by Eq.~(\ref{eq:II3}) was considered by Suzuki in order to produce path integral methods having faster asymptotic convergence. Unfortunately, the following theorem of Suzuki (see Theorem~3 of Ref.~\onlinecite{Suz91}) says that 
\begin{1}[Suzuki nonexistence theorem]
There are no finite-length splitting formulae (\ref{eq:II3}) of order $3$ or more such that the coefficients  $a_0, b_1, a_1, \ldots$ are all real and positive.
\end{1}
The Suzuki nonexistence theorem limits the asymptotic order of convergence of this type of discrete path integral methods to $2$, order of convergence that is attained for the following symmetric Trotter-Suzuki short-time approximation
\begin{equation}
\label{eq:II5}
e^{-\beta(K+V)}=e^{-\frac{1}{2}\beta V}e^{-\beta K}e^{-\frac{1}{2}\beta V}[1+{O}(\beta^3)]
\end{equation}
(or the one obtained by permuting $V$ with $K$).

The Suzuki nonexistence theorem serves to illustrate the difficulty of constructing path integral methods having asymptotic convergence better than $O(1/n^2)$. The idea of the Trotter theorem is commonly employed in the physical and chemical literature in order to generate faster integral methods starting with more general short-time approximations. The general strategy is as follows. Based upon a certain physical model, one constructs a short-time approximation $\rho_0(x,x';\beta)$ of the true density matrix. Then, one corrects upon the short-time approximation with the help of the Lie-Trotter product formula 
\begin{eqnarray} \nonumber 
\label{eq:II6}
\rho_n(x,x';\beta)=\int_{\mathbb{R}}d x_1 \ldots \int_{\mathbb{R}}d x_n\; \rho_0\left(x,x_1;\frac{\beta}{n+1}\right)\nonumber \\ \ldots \rho_0\left(x_n,x';\frac{\beta}{n+1}\right).
\end{eqnarray} 
If the short-time approximation $\rho_0(x,x';\beta)$ is ``better'' than the trapezoidal Trotter-Suzuki one, improved $n$-th order approximations to the exact density matrix may be obtained. The notion of ``better'' approximation may refer not only to the order of the short-time approximation but also to the overall quality of the approximation for finite $n$.\cite{Cep95} 

At this point, we remark that working with convergence theorems in operator norm is difficult and not particularly helpful for actual developments of better short-time approximations. Indeed, the short-time approximations are usually constructed in the configuration space as symmetric integral kernels $\rho_0(x,x';\beta)$ and many properties related to the norm operator topology are not readily available.  Therefore, it is generally more convenient to use pointwise [in the space $\mathbb{R}^2 \times [0,\infty)$ of triplets $(x,x';\beta)$] convergence theorems of the type shown by the following  theorem, which applies provided that $\rho_0(x,x';\beta)$ is symmetric. 
\begin{2}[``experimental'']
\label{th:2}
Assume that there exists the linear (automatically Hermitian) operator $T_\nu\psi$, called a convergence operator, that associates to each infinitely differentiable and compactly supported function $\psi(x)$ the square integrable function
\begin{equation}
\label{eq:II7}
(T_\nu \psi)(x) = \lim_{\beta \to 0^{+}} \frac{\int_\mathbb{R} [\rho_0(x,x';\beta) - \rho(x,x';\beta)]\psi(x') d x'}{\beta^{\nu+1}}. 
\end{equation}
Then 
\begin{eqnarray}
\label{eq:II8} \nonumber
\lim_{n\to \infty} {(n+1)^\nu}\left[\rho_n(x,x';\beta)- \rho(x,x';\beta)\right]= \\ {\beta^{\nu+1}} \int_{0}^1\left\langle x\left|e^{-\theta \beta H}T_\nu e^{-(1-\theta)\beta H}\right|x'\right\rangle 
d \theta, 
\end{eqnarray}
where $\rho_n(x,x';\beta)$ is defined by Eq.~(\ref{eq:II6}).
\end{2}
\emph{Justification.}
Let $T'_\nu(x,x';\beta)$ be defined such that
\[
\rho_0(x,x';\beta) = \rho(x,x';\beta)+ \beta^{\nu+1} T'_\nu(x,x';\beta).
\]
Lie-Trotter composing the above relation $n$ times and using the semi-group property of the exact density matrix, one argues that 
\begin{eqnarray*}\nonumber
\rho_n(x,x';\beta) = \rho(x,x';\beta) + \frac{\beta^{\nu+1}}{(n+1)^{\nu+1}}\sum_{j=0}^n \int_{\mathbb{R}}\!d x_1\! \int_{\mathbb{R}}\! d x_2\, \\ \times  \rho\left(x,x_1;\frac{j\beta}{n+1}\right) T'_\nu \left(x_1,x_2;\frac{\beta}{n+1}\right) \\ \times \rho\left(x_2,x';\frac{(n-j)\beta}{n+1}\right)  + O(1/n^{\nu+1}).\nonumber 
\end{eqnarray*}
In the limit $n \to \infty$, one uses Eq.~(\ref{eq:II7}) to cast the previous equation into
\begin{eqnarray*}\nonumber 
\lim_{n\to \infty} {(n+1)^\nu}\left[\rho_n(x,x';\beta)- \rho(x,x';\beta)\right]= \lim_{n\to \infty} \Bigg\{ \frac{\beta^{\nu+1}}{n+1} \\  \times  \sum_{j=0}^n \int_{\mathbb{R}}\!d x_1  \rho\left(x,x_1;\frac{j\beta}{n+1}\right) \left(T_\nu  \rho\right)\left(x_1,x';\frac{(n-j)\beta}{n+1}\right)  \Bigg\}.
\end{eqnarray*}
In the formula above, the operator $T_\nu$ acts upon the density matrix to the right through the first variable. Finally, one notices that in the same limit $n \to \infty$, the Riemann sum  transforms into an integral over the interval $[0,1]$, so that
\begin{eqnarray*}\nonumber &&
\lim_{n\to \infty} {(n+1)^\nu}\left[\rho_n(x,x';\beta)- \rho(x,x';\beta)\right]= \beta^{\nu+1} \\ && \times \int_0^1 d\theta \int_{\mathbb{R}}\!d x_1  \rho\left(x,x_1;\theta \beta\right) \left(T_\nu  \rho\right)\left(x_1,x';(1-\theta)\beta\right).
\end{eqnarray*}
Of course, Eq.~(\ref{eq:II8}) is nothing else than the above identity in  Dirac's bra-ket notation. \hspace{\stretch{1}}$\Box$

\emph{Observation.}  It is needless to say that the above justification is not a proof, nor is the hypothesis of the theorem completely stated.  In keeping with the scope of the  paper (and with the level of mathematical knowledge of the author), I only provide the basic reasons why the theorem must hold. The main effort of the present work is toward justifying the need for the theorems presented. The hope is that the mathematician will find the theorems interesting and worth investigating. However, all results deduced from this theorem, including the convergence constant for the trapezoidal Trotter path integral method (see Appendix~B), are verified by numerical simulations. 

Theorem~\ref{th:2} facilitates the construction of more accurate short-time approximations because it provides the exact convergence constant of the respective path integral method in coordinate representation. In general, for a given order $\nu$, one would like to design  short-time approximations $\rho_0(x,x';\beta)$ that minimize (as absolute value) the  convergence constant. In the ideal situation that the convergence constant is canceled, the order of convergence increases by one. In Section~III, we shall  use  Theorem~\ref{th:2}  to derive the set of equations that must be satisfied by the  short-time approximations of a given order $\nu$. 

\subsection{A general class of short-time approximations}

To make optimal use of Theorem~\ref{th:2}, we need to devise systematic ways of constructing symmetric and positive short-time approximations $\rho_0(x,x';\beta)$ for any   order $\nu$. The positivity of  the short-time approximation $\rho_0(x,x';\beta)$  is  necessary in order to avoid the appearance of the sign problem in the Monte Carlo simulations. Development of such systematic ways has been previously attempted by Suzuki\cite{Suz95} as well as by Makri and Miller,\cite{Mak89} among others.\cite{Rae83, Tak84, Dro98}  Unfortunately, all short-time approximations constructed so far involve derivatives of the potential $V(x)$, derivatives that are either  considered explicitly or introduced through the utilization of  commutators involving the kinetic and potential operators. In fact, the higher the convergence order, the higher the order of the derivatives that are necessary. For this reason, except for the Takahashi-Imada  approximation,\cite{Tak84} such approaches have  enjoyed only limited use. As discussed in the introduction, direct short-time high-temperature approximations based solely on the use of the potential function are more desirable.

In this subsection, I present an alternative approach to constructing  direct short-time approximations, approach that is related to the random series representation of the Brownian motion.\cite{Pre02} Evidence that will be presented in the subsequent sections supports the claim that the approach is general enough to accommodate any arbitrary convergence order $\nu$. In this work, unless otherwise specified,  $a_0, a_1, \ldots $ denotes an infinite sequence of independent identically distributed (i.i.d.)  standard normal variables. The  short-time approximations are constructed by replacing the Brownian motion in the Feynman-Kac formula with the finite dimensional Gaussian process 
\begin{equation}
\label{eq:II9}
\tilde{B}_u = \sum_{k = 0}^q a_k \tilde{\Lambda}_k(u).
\end{equation}
The continuous and piecewise smooth functions $\{\tilde{\Lambda}_k(u); 0\leq k \leq q \}$  are required to satisfy the following relations:
\begin{equation}
\label{eq:II10}
\left\{
\begin{array}{l l}
\tilde{\Lambda}_0(0) = 0, \ \tilde{\Lambda}_0(1) = 1, &   \text{and} \\
\tilde{\Lambda}_k(0) = \tilde{\Lambda}_k(1) = 0, &  \text{for} \ 1\leq k \leq q.
\end{array}
\right.
\end{equation}
  The general expression of the short-time approximations we study in the present paper is 
\begin{eqnarray}
\label{eq:II11} \nonumber
\rho_0(x,x';\beta) = \rho_{fp}(x,x';\beta) \int_{\mathbb{R}} d \mu(a_1) \cdots \int_{\mathbb{R}} d \mu(a_q) \\ \times  \exp\left\{-\beta \int_0^1 V\left[x_r(u) + \sigma \sum_{k = 1}^q a_k \tilde{\Lambda}_k(u)\right]du\right\},
\end{eqnarray}
where
\[
d\mu (a_k)= (2\pi)^{-1/2} \exp\left( - a_k^2/2 \right) d a_k
\]
and where 
\[x_r(u) = x + (x'-x)\tilde{\Lambda}_0(u)\]
is a reference path connecting the points $x$ and $x'$. 

A second condition we enforce on the system of functions $\{\tilde{\Lambda}_k(u); 0\leq k \leq q\}$ is that
\begin{equation}
\label{eq:II10b}
\tilde{\Lambda}_0(u)+\tilde{\Lambda}_0(1-u) = 1
\end{equation}
and that the finite dimensional Gaussian process $\sum_{k = 1}^q a_k \tilde{\Lambda}_k(u)$ is invariant under the transformation $u' = 1-u$. That is, we require that 
\begin{equation}
\label{eq:II11a}
\tilde{B}_u^0 = \sum_{k = 1}^q a_k \tilde{\Lambda}_k(u) \stackrel{d}{=} \sum_{k = 1}^q a_k \tilde{\Lambda}_k(1-u)=\tilde{B}_{1-u}^0.
\end{equation}
The property described by Eq.~(\ref{eq:II11a}) is  analogous to  the time symmetry of the standard Brownian bridge $B_u^0$, which is the fact that $\{B_{1-u}^0, 0\leq u \leq 1\}$ is also a Brownian bridge and is equal in distribution to $\{B_{u}^0, 0\leq u \leq 1\}$. 
As a direct consequence of Eqs.~(\ref{eq:II10b}) and~(\ref{eq:II11a}), the  short-time approximation $\rho_0(x,x';\beta)$ given by Eq.~(\ref{eq:II11}) is symmetric under the permutation of the variables $x$ and $x'$. This can be verified by performing the substitution $u' = 1- u$ in Eq.~(\ref{eq:II11}). The time symmetry of the finite Gaussian process $\sum_{k = 1}^q a_k \tilde{\Lambda}_k(u)$ can be enforced, for example, by restricting the functions $\{\tilde{\Lambda}_k(u); 1\leq k \leq q\}$ to the class of  symmetric and antisymmetric functions. 

In this general setting, given a fixed integer $q$, Theorem~\ref{th:2} suggests that the functions $\{\tilde{\Lambda}_k(u); 0\leq k \leq q \}$  should be chosen such that the order of convergence be maximized. We shall show in the next section that the system of functional equations controlling the order of convergence is \emph{independent} of  the potential $V(x)$. This system of equations does not uniquely determine  the  functions $\{\tilde{\Lambda}_k(u); 0\leq k \leq q \}$. For instance, it is a trivial matter to show that the short-time approximation given by Eq.~(\ref{eq:II11}) is invariant under a linear orthogonal transformation of the functions $\{\tilde{\Lambda}_k(u); 1\leq k \leq q \}$. 

A consequence of the constraint given by Eq.~(\ref{eq:II10}) is the fact that the distributions of the end points $B_1$ and $\tilde{B}_1$ are identical and equal to that of the variable $a_0$. In order to reproduce in a better way the properties of the Brownian motion, we may also require (but it is not necessary) that the pairs of Gaussian variables $(B_1, M_1)$ and $(\tilde{B}_1, \tilde{M}_1)$ have equal joint distribution. Here, $M_1$ and $\tilde{M}_1$ are the so-called path centroids\cite{Fey65a} (first moments of the Brownian motion and its short-time approximation) and are defined by the equations
\[
M_1 = \int_0^1 B_u du \qquad \text{and} \qquad \tilde{M}_1 = \int_0^1 \tilde{B}_u du,
\]
respectively. To find the class of  short-time approximations for which this condition is ``built in,'' consider $\lambda_0(u) = 1$ and $\lambda_1(u) =\sqrt{3}(1 - 2u) $, the first two normalized Legendre polynomials on the interval $[0,1]$. Let $\{\lambda_k(u)\}_{k \geq 2}$ be a set of functions which together with the first two Legendre polynomials make up an orthonormal set on $[0,1]$. The Ito-Nisio theorem (see Theorem~\ref{th:6} of Appendix~A) says that 
\[
B_u \stackrel{d}{=} a_0 u + a_1 \sqrt{3} u(1-u)+ \sum_{k = 2}^\infty a_k \Lambda_k(u),
\]
where 
\[
\Lambda_k(u) = \int_0^u \lambda_k(\tau) d\tau. 
\]
Let us notice that if $k \geq 2$, then $\Lambda_k(1) = 0$ [by the orthogonality of $\lambda_k(u)$ on $1$] and 
\[
\int_0^1 \Lambda_k(u)du = \Lambda_k(1) - \int_0^1 \lambda_k(u) u du = 0
\]
 [by the orthogonality of $\lambda_k(u)$ on $u$].
Therefore, $B_1 = a_0$ and
\[
M_1 = \frac{1}{2}a_0 + \frac{\sqrt{3}}{6} a_1  
\]
depend solely on the variables $a_0$ and $a_1$. A little thought shows that we can build in the correct joint distribution of the end point and the path centroid by further restricting the class of functions $\{\tilde{\Lambda}_k(u); 0\leq k \leq q \}$  to those satisfying the constraints
\begin{equation}
\label{eq:II15}
\left\{
\begin{array}{l l}
\int_0^1 \tilde{\Lambda}_k(u)du = 1/2, & \text{for} \ k = 0, \\
\int_0^1 \tilde{\Lambda}_k(u)du = \sqrt{3}/6, & \text{for} \ k = 1, \ \text{and} \\
\int_0^1 \tilde{\Lambda}_k(u)du= 0,& \text{for} \ 2 \leq k \leq q. \\
\end{array}
\right.
\end{equation}

Until now, we have assumed that the path averages of the type 
\[
\int_0^1 V\left[x_r(u) + \sigma \sum_{k = 1}^q a_k \tilde{\Lambda}_k(u)\right]du
\]
are evaluated exactly. For practical applications, one also needs to devise a minimalist quadrature scheme specified by some points $0 \leq u_1 < u_2 < \ldots < u_{n_q} \leq 1$ and nonnegative weights $w_1, w_2, \ldots, w_{n_q}$ such that the discrete short-time approximation 
\begin{eqnarray}
\label{eq:II16} \nonumber
\rho_0(x,x';\beta) = \rho_{fp}(x,x';\beta) \int_{\mathbb{R}} d \mu(a_1) \cdots \int_{\mathbb{R}} d \mu(a_q) \\ \times  \exp\left\{-\beta \sum_{i = 1}^{n_q}w_i V\left[x_r(u_i) + \sigma \sum_{k = 1}^q a_k \tilde{\Lambda}_k(u_i)\right]\right\}
\end{eqnarray}
has the desired convergence order. In the case of the discrete approximations, the set of quadrature points and weights as well as the values of the functions $\tilde{\Lambda}_k(u)$ at the quadrature points are fitting parameters.  For the reason of ensuring time symmetry of the discrete formula, the quadrature scheme is required to be symmetric, i.e., the sequences  $u_1, u_2 - u_1,  \ldots,  1 - u_{n_q} $ and  $w_1, w_2, \ldots, w_{n_q}$ must be palindromic.

\section{Power series expansion for imaginary-time propagated wavefunctions} 

In this section, we shall derive the system of functional equations that must be satisfied by the functions $\tilde{\Lambda}_k(u)$ appearing in Eq.~(\ref{eq:II9}) in order for the associated  short-time approximation to have a convergence order $\nu$. To settle some terminology related to the utilization of the term ``short-time,'' we interpret the parameter $\beta$ as a time variable (physically, $\hbar \beta$ has dimension of time) so that the density matrix $\rho(x,x';\beta)$ constitutes the time-dependent Green's function of a diffusion equation, or imaginary-time Schr{\"o}dinger equation.   As Theorem~\ref{th:2} illustrates, it is necessary to establish the power series expansion of the imaginary-time propagated wavefunctions for the exact and the approximate propagators, respectively. I warn the reader that the power series derived in the present section are only a bookkeeping device for derivatives against $\beta$ and are not required to converge to the actual imaginary-time propagated solutions. Moreover, the potential $V(x)$ and its derivatives are required to have finite Gaussian transforms. Actually, we require that
\begin{equation}
\label{eq:0}
\frac{1}{\sqrt{2\pi \alpha}}\int_{\mathbb{R}} e^{-z^2/(2 \alpha)} |V^{(k)}(x + z)|^j < \infty
\end{equation}
for all $x \in \mathbb{R}$ and $\alpha > 0$, as well as for all integers $k \geq 0$ and $j \geq 1$. This condition is necessary in order to ensure that we recover the original potentials, derivatives, or products of such functions from their Gaussian transforms, in the limit that $\alpha \to 0$ (see Theorem~3 of Ref.~\onlinecite{Pre02f}). 

\subsection{The exact propagator}

The power series expansion of the propagated wavefunction
\begin{equation}
\label{eq:1}
\left\langle x \left| e^{-\beta H} \right| \psi \right\rangle = \int_\mathbb{R} \rho(x,x';\beta)\psi(x') d x'
\end{equation}
is of utmost interest for the present development. With the help of the Feynman-Kac formula [according to Eq.~(\ref{eq:B4}) of Appendix~A] and the Taylor power series expansion, one writes
\begin{eqnarray*}
\nonumber &&
\left\langle x \left| e^{-\beta H} \right| \psi \right\rangle = \mathbb{E} \left[ e^{-\beta \int_0^1 V(x+\sigma B_u)du} \psi(x + \sigma B_1) \right]\\&& = \mathbb{E} \left\{  \left[ \sum_{j = 0}^\infty \frac{1}{j!} \psi^{(j)}(x) \sigma^j (B_1)^j\right]\prod_{k = 0}^\infty e^{-\beta\frac{V^{(k)}(x)}{k!}\sigma^k M_k}\right\},
\end{eqnarray*}
where
\begin{equation}
M_k = \int_0^1 (B_u)^k du.
\end{equation}
A second Taylor expansion leads to 
\begin{eqnarray*}
\nonumber 
\left\langle x \left| e^{-\beta H} \right| \psi \right\rangle = \mathbb{E} \left\{  \left[ \sum_{j = 0}^\infty \frac{1}{j!} \psi^{(j)}(x) \sigma^j (B_1)^j\right]\right. \\ \left. \times \prod_{k = 0}^\infty \left[\sum_{j = 0}^\infty \frac{(-\beta)^j}{j!} \frac{V^{(k)}(x)^j}{(k!)^j}\sigma^{kj} \left(M_k\right)^j\right]\right\}.
\end{eqnarray*}
We now expand the product in the preceding formula and collect the coefficients corresponding to the same power of $\beta$. Remembering that $\sigma = \left(\hbar^2/m_0\right)^{1/2} \beta^{1/2}$, one argues that the powers of $\beta$ are of the form $\beta^\mu$, where $\mu$ is a non-negative half integer, i.e. an element of the set $\mathbb{N}_2 = \left\{n/2: n \in \mathbb{N}\right\}$.  For each $\mu \in \mathbb{N}_2$, $\mu > 0$, define 
\begin{equation}
\label{eq:3}
J_\mu= \left\{ (j_1, j_2, \ldots, j_{2\mu})\in \mathbb{N}^{2\mu}: \sum_{k = 1}^{2\mu}k j_k  = 2\mu \right\}.
\end{equation} 
A little thought shows that 
\begin{widetext}
\begin{eqnarray}
\label{eq:4} \nonumber &&
\left\langle x \left| e^{-\beta H} \right| \psi \right\rangle = \sum_{\mu \in \mathbb{N}_2} \beta^\mu \sum_{(j_1,\ldots,j_{2\mu})\in J_\mu} (-1)^{j_2 + \ldots + j_{2\mu}}\left(\hbar^2/m_0\right)^{\frac{j_1 + j_3 + 2j_4 + \ldots + (2\mu-2)j_{2\mu}}{2}} \\ && \times \frac{ \psi^{(j_1)}(x)V(x)^{j_2}\left[V^{(1)}(x)\right]^{j_3}\ldots \left[V^{(2\mu-2)}(x)\right]^{j_{2\mu}}} {j_1!j_2!\ldots j_{2\mu}! (2!)^{j_4}(3!)^{j_5} \ldots [(2\mu-2)!]^{j_{2\mu}} } \mathbb{E}\left[(B_1)^{j_1}(M_0)^{j_2}(M_1)^{j_3}\ldots (M_{2\mu-2})^{j_{2\mu}}\right],
\end{eqnarray}
\end{widetext}
with the convention that the term for $\mu = 0$ is $\psi(x)$.
The fact that $B_u$ is a Gaussian distributed variable of mean zero implies that if $j_1 + j_3+ 2j_4 + \ldots + (2\mu-2)j_{2\mu}$ is odd, then
\[\mathbb{E}\left[(B_1)^{j_1}(M_0)^{j_2}(M_1)^{j_3}\ldots (M_{2\mu-2})^{j_{2\mu}}\right] = 0,\]
as can be verified by induction. 
Since $j_1 + j_3+ 2j_4 + \ldots + (2\mu-2)j_{2\mu} = 2\mu - 2(j_2 + \dots + j_{2\mu})$, one sees that $j_1 + j_3+ 2j_4 + \ldots + (2\mu-2)j_{2\mu}$ is odd if and only if $2\mu$ is an odd integer. Thus, the sum in Eq.~(\ref{eq:4}) can be restricted to the numbers $\mu \in \mathbb{N}_2$ for which $2\mu$ is even i.e., the sum can be restricted to the set of natural numbers $\mathbb{N}$. Therefore,
\begin{widetext}
\begin{eqnarray} 
\label{eq:5} \nonumber &&
\int_\mathbb{R} \rho(x,x';\beta)\psi(x')dx' = \sum_{\mu =0}^\infty \beta^\mu \sum_{(j_1,\ldots,j_{2\mu})\in J_\mu} (-1)^{j_2 + \ldots + j_{2\mu}}\left(\hbar^2/m_0\right)^{\mu - (j_2 + \dots + j_{2\mu})} \\ && \times \frac{ \psi^{(j_1)}(x)V(x)^{j_2}\left[V^{(1)}(x)\right]^{j_3}\ldots \left[V^{(2\mu-2)}(x)\right]^{j_{2\mu}}} {j_1!j_2!\ldots j_{2\mu}! (2!)^{j_4}(3!)^{j_5} \ldots [(2\mu-2)!]^{j_{2\mu}} } \mathbb{E}\left[({B}_1)^{j_1}(M_0)^{j_2}({M}_1)^{j_3}\ldots ({M}_{2\mu-2})^{j_{2\mu}}\right].
\end{eqnarray}
\end{widetext}

\emph{Observation.} The power series expansion for the imaginary-time propagated wavefunction can also be derived by expanding the operator $e^{-\beta H}$ in a power series. One obtains
\begin{eqnarray}
\label{eq:5a} \nonumber &&
\left\langle x \left| e^{-\beta H} \right| \psi \right\rangle = \sum_{\mu = 0}^\infty \frac{1}{\mu!} \left \langle x \left|\left(-\beta H\right)^\mu \right| \psi \right\rangle  \\ && = \sum_{\mu = 0}^\infty \beta^\mu \frac{1}{\mu!} \left\{ \left[\frac{\hbar^2}{2 m_0}\frac{d^2}{dx^2} - V(x) \right]^\mu \psi(x) \right\}.
\end{eqnarray}
Of course, the terms of the two series given by Eqs.~(\ref{eq:5}) and (\ref{eq:5a}) are equal.  However, as we shall see in the following subsection, Eq.~(\ref{eq:5}) applies in an almost unchanged form for all short-time approximations defined by Eqs.~(\ref{eq:II11}) and (\ref{eq:II16}). In contrast, there might be no formal analogue of Eq.~(\ref{eq:5a}) for such short-time approximations.

\subsection{The approximate propagator and the identities controlling its order of convergence}
The only property used for the derivation of the power series expansion of the exact propagator was the fact that the Brownian motion is a Gaussian process. Since the  approximation to the Brownian motion given by Eq.~(\ref{eq:II9}) is also a Gaussian process, Eq.~(\ref{eq:5}) remains true for the approximate propagator, too. Therefore, 
\begin{widetext}
\begin{eqnarray}
\label{eq:6} \nonumber &&
\int_\mathbb{R} \rho_0(x,x';\beta)\psi(x')dx' = \sum_{\mu =0}^\infty \beta^\mu \sum_{(j_1,\ldots,j_{2\mu})\in J_\mu} (-1)^{j_2 + \ldots + j_{2\mu}}\left(\hbar^2/m_0\right)^{\mu - (j_2 + \dots + j_{2\mu})} \\&&  \times \frac{ \psi^{(j_1)}(x)V(x)^{j_2}\left[V^{(1)}(x)\right]^{j_3}\ldots \left[V^{(2\mu-2)}(x)\right]^{j_{2\mu}}} {j_1!j_2!\ldots j_{2\mu}! (2!)^{j_4}(3!)^{j_5} \ldots [(2\mu-2)!]^{j_{2\mu}} } \mathbb{E}\left[(\tilde{B}_1)^{j_1}(\tilde{M}_0)^{j_2} (\tilde{M}_1)^{j_3}\ldots (\tilde{M}_{2\mu-2})^{j_{2\mu}}\right],
\end{eqnarray}
\end{widetext}
where 
\begin{equation}
\label{eq:7}
\tilde{M}_k = \int_0^1 \left(\tilde{B}_u\right)^k d u.
\end{equation}
If the discrete  short-time approximation given by Eq.~(\ref{eq:II16}) is employed, then Eq.~(\ref{eq:6}) remains true provided that $\tilde{M}_k$ is redefined to be
\begin{equation}
\label{eq:8}
\tilde{M}_k = \sum_{i = 1}^{n_q}w_i \left(\tilde{B}_{u_i}\right)^k.
\end{equation}

Theorem~\ref{th:2} immediately implies the following statement. 
\begin{4}
\label{th:4}
A short-time approximation of the types given by Eq.~(\ref{eq:II9}) or Eq.~(\ref{eq:II16}) has convergence order $\nu$ if and only if
\begin{eqnarray}
\label{eq:9}&& \nonumber
 \mathbb{E}\left[({B}_1)^{j_1}(M_0)^{j_2}({M}_1)^{j_3}\ldots ({M}_{2\mu-2})^{j_{2\mu}}\right] \\ && = \mathbb{E}\left[(\tilde{B}_1)^{j_1}(\tilde{M}_0)^{j_2} (\tilde{M}_1)^{j_3}\ldots (\tilde{M}_{2\mu-2})^{j_{2\mu}}\right]
\end{eqnarray}
for all $2\mu$-tuples of non-negative integers $(j_1, j_2, \ldots, j_{2\mu})$ such that 
\[
\sum_{k = 1}^{2\mu} k j_k = 2\mu 
\]
and $1 \leq \mu \leq \nu$.
\end{4}

The general problem that one would like to solve using the theory developed so far is the following. Given a convergence order $\nu$, is there a finite system of functions $\tilde{\Lambda}_k(u)$ such that the corresponding  short-time approximation has order $\nu$? If the answer is yes,  what is the minimal number $q$ of functions $\tilde{\Lambda}_k(u)$ necessary to achieve the respective convergence order? Then, what is the minimal number of quadrature points such that a discrete  short-time approximation has convergence order $\nu$? The relevance of the questions asked in the current paragraph will be further clarified in Section~VI, where we  analyze the problem of minimizing the statistical noise for real-time propagators. 

\section{Examples of short-time approximations having convergence order $3$ or $4$}
In this section, I try to present evidence in support of the idea that the system of equations appearing in Theorem~\ref{th:4} for a given order $\nu$ is always satisfied by some finite system of functions $\tilde{\Lambda}_k(u)$. I do this by computing explicit numerical solutions for the convergence orders $3$ and $4$. As apparent from Table~\ref{Tab:I}, the number of equations that need to be verified for a given order $\nu$ increases rapidly with $\nu$. In fact, the number of elements of $J_\mu$ is the number of distinct partitions of $2\mu$. With the help of the Hardy-Ramanujan asymptotic formula,\cite{Har99} one deduces that the number of equations that need to be verified for a given order $\nu$  behaves asymptotically as
\begin{equation*}
 \sum_{\mu = 1}^\nu \frac{1}{8\mu\sqrt{3}} e^{\pi \sqrt{4\mu/3}}.
\end{equation*} 
Therefore, the ``by hand'' approach utilized in the present section is bound to fail even for slightly larger convergence orders. By use of computers, one may hope to obtain solutions for moderately large convergence orders. However, I believe future work on the problem may reveal better strategies for the computation of short-time approximations of high convergence orders. 

The two short-time approximations constructed in the present section are called reweighted short-time approximations.\cite{Pre03} The defining features are  the equality $\tilde{\Lambda}_0(u) = u$ and the fact that the functions $\{\tilde{\Lambda}_k(u); 1 \leq k \leq q\}$ appearing in Eq.~(\ref{eq:II9}) are required to satisfy the constraint
\begin{equation}
\label{eq:II10a}
\sum_{k = 1}^q \tilde{\Lambda}_k(u)^2= u(1-u). 
\end{equation}
The last equation  stems from the condition that  the Gaussian variables $B_u$ and $\tilde{B}_u$ have equal variances for each $u \in [0,1]$ (equal weights). As we shall see, if this constraint is imposed, most of the functional equations for convergence orders $3$ and $4$ are automatically satisfied. However,  the number of remaining equations still scales exponentially and, for higher convergence orders, the constraint given by Eq.~(\ref{eq:II10a}) may actually become a nuisance.

One additional feature of the reweighted  short-time approximations  stems from the relation $\tilde{\Lambda}_0(u) = u$ and facilitates the numerical implementation of the associated Lie-Trotter product formula given by Eq.~(\ref{eq:II6}). The following generalization of a result of Predescu and Doll (see Theorem~2 of Ref.~\onlinecite{Pre02b}) is straightforward to prove. 

Assume $n$ is of the form $n = 2^k -1$ and let $\{a_{l,j}; 1\leq l \leq k, 1 \leq j \leq 2^{l-1}\}$ and $\{b_{l,j}; 1\leq l \leq q, 1\leq  j \leq 2^{k}\}$ be two independent sets of i.i.d. standard normal variables. Let $\{F_{l,j}(u); l\geq 1, 1\leq j \leq 2^{l-1}\}$ be the system of Schauder functions\cite{McK69} on the interval $[0,1]$. The Schauder functions can be generated by translations and dilatations as follows. Let 
$F_{1,1}(u): \mathbb{R} \to \mathbb{R}$  be defined by 
\begin{equation}
\label{eq:1.89}
F_{1,1}(u) = \left\{\begin{array}{cc} u,& u \in [0, 1/2],\\ 1-u,& u \in (1/2, 1],\\ 0, &\text{elsewhere}. \end{array}\right.
\end{equation}
Then,
\begin{equation}
\label{eq:1.91}
F_{l,j}(u)= 2^{-(l-1)/2} F_{1,1}(2^{l-1}u - j + 1),
\end{equation}
for all $l\geq 1$ and $1\leq j \leq 2^{l-1}$. Extend the functions $\{\tilde{\Lambda}_l(u); 1 \leq l \leq q\}$ outside the interval $[0,1]$ by setting them to zero [the same way the first Schauder function $F_{1,1}(u)$ was extended to the whole real axis in Eq.~(\ref{eq:1.89})] and define
 \begin{equation}
\label{eq:1.92}
G_{l,j}(u) = 2^{-k/2}\tilde{\Lambda}_l(2^k u - j + 1),
\end{equation}
for $1\leq l \leq q$ and $1 \leq j \leq 2^k$. 

In these conditions, the following theorem holds. 
\begin{3}
\begin{widetext}
With the convention that $a_{l, 2^{l-1}+1} = 0$ and $b_{l,2^k+1} = 0$ for all $l\in \overline{1,k}$, we have
\begin{eqnarray}
\label{eq:II17} \nonumber
\frac{\rho_n(x, x' ;\beta)}{\rho_{fp}(x, x' 
;\beta)}&=&\int_{\mathbb{R}} d a_{1,1}\ldots \int_{\mathbb{R}} d a_{k,2^{k-1}}  \left( 2\pi \right)^{-n/2}  \exp\left({-\frac{1}{2}\sum_{l=1}^{k}\sum_{j=1}^{2^{l-1}}  a_{l,j}^2}\right) \\ & \times & \nonumber
\int_{\mathbb{R}} d b_{1,1}\ldots \int_{\mathbb{R}} d b_{q,2^k}  \left( 2\pi \right)^{-(n+1)q/2}  \exp\left({-\frac{1}{2}\sum_{l=1}^{q}\sum_{j=1}^{2^{k}}  b_{l,j}^2}\right)
\\& \times & \exp\left\{-\beta \int_0^1 V\left[x_r(u)+\sigma \sum_{l=1}^{k} a_{l,[2^{l-1} u]+1} \;{F}_{l,[2^{l-1} u]+1}(u) \right. \right. \\ && \left. \left.+ \sigma \sum_{l=1}^{q} b_{l,[2^k u]+1} \;G_{l,[2^k u]+1}(u)\right]d u\right\}, \nonumber
\end{eqnarray}
\end{widetext}
where $[2^{l-1} u]$ and $[2^k u]$ are the integer parts of $2^{l-1} u$ and $2^k u$, respectively.
\end{3}

The reader can easily verify that Eq.~(\ref{eq:II17}) is a so-called reweighted L\'evy-Ciesielski path integral technique, as defined in Ref.~\onlinecite{Pre03}. It has been argued\cite{Pre02b} that this representation is more advantageous  than the direct expression of $\rho_n(x,x';\beta)$ that is obtained from the Lie-Trotter product formula, for practical implementations. The expression obtained by Lie-Trotter composing the discrete version of $\rho_0(x,x';\beta)$ given be Eq.~(\ref{eq:II16}) can also be put in the form of Eq.~(\ref{eq:II17}). However, the  one-dimensional integral at exponent is replaced by a quadrature sum. The quadrature scheme is specified by the $n_q 2^k$ (not necessarily  different) quadrature points 
\begin{equation}
\label{eq:II18}
u'_{i,j} = 2^{-k}(u_i + j - 1), \quad   1 \leq i \leq n_q, \ 1 \leq j \leq 2^k
 \end{equation}
and the corresponding weights 
\begin{equation}
\label{eq:II19}
w'_{i,j} = 2^{-k}w_i.
\end{equation}
The new quadrature points $u'_{i,j}$  are obtained by translations and dilatations (more precisely, contractions) of the original quadrature points $u_i$.

\begin{table}[!tbh]
%\begin{table}
\caption{Indexes of the equations that need to be verified for various values of $\mu$. Shown are the non-zero components of these indexes. \label{Tab:I}
}
\begin{tabular}{ | c | c |}
   \hline \hline
$\mu = 1$ & $j_2 = 1$ \\
          & $j_1 = 2$ \\
\hline    
					& \\
					& $j_4 = 1$\\
          & $j_3 = 1$, $j_1 = 1$\\
$\mu = 2$ & $j_2 = 2$\\
          & $j_2 = 1$, $j_1 = 2$\\
          & $j_1 = 4$\\
          & \\
\hline
	       	& \\
          & $j_6 = 1$\\
					& $j_5 = 1$, $j_1 = 1$\\
					& $j_4 = 1$, $j_2 = 1$\\
					& $j_4 = 1$, $j_1 = 2$\\
        	& $j_3 = 2$\\
$\mu = 3$ & $j_3 = 1$, $j_2 = 1$, $j_1 = 1$\\
					& $j_3 = 1$, $j_1 = 3$\\
					& $j_2 = 3$\\
					& $j_2 = 2$, $j_1 = 2$\\
					& $j_2 = 1$, $j_1 = 4$\\
					& $j_1 = 6$\\
					& \\
\hline \hline
\end{tabular}
\begin{tabular}{|c|c| }
   \hline \hline
          & $j_8 = 1$ \\
          & $j_7 = 1$, $j_1 = 1$ \\
          & $j_6 = 1$, $j_2 = 1$ \\
          & $j_6 = 1$, $j_1 = 2$ \\
          & $j_5 = 1$, $j_3 = 1$ \\
          & $j_5 = 1$, $j_2 = 1$, $j_1 = 1$\\
          & $j_5 = 1$, $j_1 = 3$ \\
          & $j_4 = 2$\\
					& $j_4 = 1$, $j_3 = 1$, $j_1 = 1$\\
					& $j_4 = 1$, $j_2 = 2$\\
$\mu = 4$	& $j_4 = 1$, $j_2 = 1$, $j_1 = 2$\\
					& $j_4 = 1$, $j_1 = 4$\\
					& $j_3 = 2$, $j_2 = 1$\\
					& $j_3 = 2$, $j_1 = 2$\\
					& $j_3 = 1$, $j_2 = 2$, $j_1 = 1$\\
					& $j_3 = 1$, $j_2 = 1$, $j_1 = 3$\\
					& $j_3 = 1$, $j_1 = 5$\\
					& $j_2 = 4$\\
					& $j_2 = 3$, $j_1 = 2$\\
					& $j_2 = 2$, $j_1 = 4$\\
					& $j_2 = 1$, $j_1 = 6$\\
					& $j_1 = 8$\\
\hline \hline
\end{tabular}
%\end{table}
\end{table}

\subsection{Reweighted short-time approximation having convergence order $3$}

The  equations that the functions $\tilde{\Lambda}_k(u)$ must satisfy in order to generate a reweighted short-time approximation of order $3$ are those of the type shown by Eq.~(\ref{eq:9}) for the indexes $(j_1, j_2, \ldots, j_{2\mu})$ presented in Table~\ref{Tab:I}, with $\mu = 1$, $2$, and $3$. For a better understanding, we mention that in Table~\ref{Tab:I} we only present the non-zero components of a given index $(j_1, j_2, \ldots, j_{2\mu})$.  There are a total of $2 + 5 + 11 = 18$ equations that should be verified. However, given the special form of the reweighted finite-dimensional approximation to the Brownian motion,  most of these equations are automatically satisfied. As such, the equations for which the only non-zero components are $j_1$ and $j_2$ are verified by all reweighted short-time approximations. The discrete versions satisfy the respective equations provided that 
\[
\sum_{i = 1}^{n_q} w_i = 1. 
\]
One actually checks that all equations for $\mu = 2$ as well as all equations for $\mu = 3$, except for the one specified by $j_2 = 2$, are automatically satisfied. The discrete version verifies these equations provided that the quadrature scheme is capable of integrating exactly all polynomials $1$, $u$, and $u^2$. For example, let us consider the equation specified by $j_6 = 1$. We have
\[
\mathbb{E} \left[\sum_{i = 1}^{n_q} w_i (\tilde{B}_{u_i})^4\right] = \sum_{i = 1}^{n_q}w_i \mathbb{E} \left[(\tilde{B}_{u_i})^4 \right]= 3\sum_{i = 1}^{n_q}w_i u_i^2. 
\]
By Eq.~(\ref{eq:9}) as specialized for $j_6 = 1$, the above value should equal [see Eq.~(\ref{eq:B2}) of Appendix~A] 
\[
\mathbb{E}\left[\int_0^1 (B_{u})^4 du\right] = \int_0^1 \mathbb{E}\left[(B_{u})^4\right] du = 3\int_0^1 u^2 du.
\]
This shows that the quadrature technique must integrate exactly the polynomial $u^2$. 

We now turn our attention to the remaining equation defined by $j_3 = 2$. 
One computes
\begin{equation}
\label{eq:10}
\mathbb{E}\left[\int_0^1 \tilde{B}_{u} du\right]^2 = \left[\int_0^1 u du \right]^2 + \sum_{k = 1}^q \left[ \int_0^1 \tilde{\Lambda}_k (u)du \right]^2,
\end{equation}
which should equal 
\begin{equation}
\label{eq:11}
\mathbb{E}\left[\int_0^1 B_{u} du\right]^2 = \left[\int_0^1 u du \right]^2 + 3 \left[ \int_0^1 u(1-u)du\right]^2. 
\end{equation}
To compute the expected value of the square of the first moment of the Brownian motion, write the Brownian motion as a random series constructed via the Ito-Nisio theorem from the Legendre orthogonal polynomials on the interval $[0,1]$. Then, as discussed in the preceding section, 
\[
\int_0^1 B_u du = a_0 \int_0^1 u du + \sqrt{3} a_1 \int_0^1 u(1-u) du
\]
and Eq.~(\ref{eq:11}) follows. From Eqs.~(\ref{eq:10}) and (\ref{eq:11}), one easily obtains the identity
\[
\sum_{k = 1}^q \left[ \int_0^1 \tilde{\Lambda}_k (u)du \right]^2 = \frac{1}{12}. 
\]
A similar relation can be deduced for the discrete version but with the integrals replaced by the corresponding quadrature sums. 

We can summarize the findings of the present subsection into the following proposition. 

\begin{Pr1}
A reweighted short-time approximation has order $3$ if and only if 
\begin{equation}
\label{eq:12}
\sum_{k = 1}^q \left[ \int_0^1 \tilde{\Lambda}_k (u)du \right]^2 = \frac{1}{12}. 
\end{equation}
A discrete reweighted short-time approximation has order $3$ provided that the associated quadrature scheme integrates exactly all polynomials of degree at most $2$ and provided that 
\begin{equation}
\label{eq:13}
\sum_{k = 1}^q \left[ \sum_{i = 1}^{n_q}w_i \tilde{\Lambda}_k (u_i) \right]^2 = \frac{1}{12}. 
\end{equation}
\end{Pr1}

We conclude the present subsection by constructing a minimalist reweighted short-time approximation having convergence order $3$. Because of the identity (\ref{eq:II10a}), the minimal number $q$ of functions $\tilde{\Lambda}_k(u)$ capable of satisfying Eq.~(\ref{eq:12}) is $2$. Indeed, if $q$ = 1, then $\tilde{\Lambda}_1(u)= [u(1-u)]^{1/2}$ and 
\[
\left[ \int_0^1 \tilde{\Lambda}_1 (u)du \right]^2 = \pi^2/64 \neq 1/12.
\]
We now try a set of two functions of the form 
\begin{eqnarray}
\label{eq:14}
\left\{\begin{array}{c} 
\tilde{\Lambda}_1(u) = \sqrt{u(1-u)} \cos[\alpha(u-0.5)], \\ 
 \tilde{\Lambda}_2(u) = \sqrt{u(1-u)} \sin[\alpha(u-0.5)]. 
\end{array}\right.
\end{eqnarray}
The functions $\tilde{\Lambda}_1(u)$ and $\tilde{\Lambda}_2(u)$ are orthogonal because the first is symmetric under the transformation $u'= 1- u$, whereas the second is antisymmetric. 
The constant $\alpha$ is then determined by Eq.~(\ref{eq:12}) and has been evaluated with the help of the Levenberg-Marquardt algorithm, as implemented in Mathcad.\cite{Mathcad} 
The solution has the approximate value
\begin{equation}
\label{eq:15}
\alpha \approx 3.056620471.
\end{equation}

To design a minimalist discrete short-time approximation of order $3$, we consider an arbitrary symmetric quadrature rule on the interval $[0,1]$ that integrates exactly all polynomials of degree less or equal to $2$. Then, we find the value of $\alpha$ that satisfies Eq.~(\ref{eq:13}) for the chosen quadrature technique. It is not difficult to argue that the minimal number of quadrature points in the open interval $(0,0.5)$ must be $1$. The reason is that the values of the functions $\tilde{\Lambda}_1(u)$ and $\tilde{\Lambda}_2(u)$   at the points $u = 0$ and $u = 0.5$ do not depend upon the parameter $\alpha$. Thus, Eq.~(\ref{eq:13}) cannot be satisfied if there are no quadrature points located inside the open interval $(0,0.5)$.

The quadrature rule is taken to be the $2$-point Gauss-Legendre rule on the interval $[0,1]$, quadrature rule that integrates exactly all polynomials of degree less or equal to $3$. 
The appropriate value for the parameter $\alpha$ is then determined from Eq.~(\ref{eq:13}) and is found to be
\begin{equation}
\label{eq:15a}
\alpha \approx 2.720699046.
\end{equation}
 The quadrature scheme is given in Table~\ref{Tab:II}, for ease of reference. 
\begingroup
\begin{table}[!htbp]
\caption{
\label{Tab:II}
Quadrature points and weights for the minimalist discrete short-time approximation of order $3$. The points and weights are those for the $2$-point Gauss-Legendre rule on the interval $[0,1]$.}
\begin{tabular}{|c |c |c |}
\hline
$i$ & 1 & 2   \\ 
\hline \hline
$u_i$&0.211324865&0.788675135\\
\hline
$w_i$&0.500000000&0.500000000\\
\hline 
\end{tabular}
\end{table}
\endgroup

As shown by Eq.~(\ref{eq:II17}), the number of path variables entering the expression of $\rho_n(x,x';\beta)$ is $(q+1)n + q = 3n+2$, whereas the number of quadrature points [see Eq.~(\ref{eq:II18})] is $n_q(n+1)= 2n+2$. Thus, for large enough $n$, the ratio $(2n+2) / (3n+2)$ approaches $2/3$, value that is smaller than the one for the trapezoidal Trotter discrete path integral method. Therefore, the method described in the present paragraph has fewer numerical requirements than the trapezoidal Trotter discrete path integral method for equal numbers of path variables, yet it achieves cubic convergence for smooth enough potentials.  

\subsection{Reweighted short-time approximation having convergence order $4$}

Because the number of  equations to be verified increases significantly for the reweighted short-time approximations of order $4$, we choose to approximate the Brownian motion by the finite dimensional process 
\begin{equation}
\label{eq:16}
\tilde{B}_u \stackrel{d}{=} a_0 u + a_1 \sqrt{3} u(1-u)+ \sum_{k = 2}^q a_k \tilde{\Lambda}_k(u),
\end{equation}
where the functions $\tilde{\Lambda}_k(u)$ satisfy the equations
\[
\int_0^1 \tilde{\Lambda}_k(u) du = 0, \quad \text{for} \ 2 \leq k \leq q. 
\]
As discussed in Section~II.B, in this case the variables $(\tilde{B}_0, \tilde{M}_0, \tilde{M}_1)$ have the same joint distribution as $(B_0, M_0, M_1)$ (notice that $\tilde{M}_0$ and $M_0$ are equal constants). This remains true of the discrete  reweighted short-time approximations provided that the quadrature scheme integrates exactly the polynomials of degree at most $2$ as well as the functions $\tilde{\Lambda}_k(u)$, for $2 \leq k \leq q$. 

Using the special form of Eq.~(\ref{eq:16}), it is not difficult to verify that all the equations in Table~\ref{Tab:I} are automatically satisfied with the exception of the one specified by $j_4 = 2$. This remains true of the discrete versions provided that the quadrature scheme integrates exactly all polynomials of degree at most $3$ as well as the functions $\tilde{\Lambda}_k(u)$, for $2 \leq k \leq q$. For the sake of an example, let us consider the equation specified by  $j_5 = 1, j_3 = 1$, which is the most difficult to verify. 
I leave it for the reader to argue that in general
\begin{eqnarray}\nonumber
\label{eq:17}
\mathbb{E}\left(\sum_{i_1, i_2, i_3, i_4} a_{i_1}a_{i_2}a_{i_3}a_{i_4}M_{i_1,i_2, i_3, i_4}\right)\\ = \sum_{i,j} \left(M_{i,i,j,j} + M_{i,j,i,j} + M_{i,j,j,i}\right).
\end{eqnarray}
 Using Eq.~(\ref{eq:17}), one computes
\begin{eqnarray*}&&
\mathbb{E}\left(\int_0^1 \tilde{B}_u du \int_0^1 \tilde{B}_u^3 du\right)= 3 \sum_{i,j = 0}^q \left[\int_0^1 \tilde{\Lambda}_i(u)du\right. \\ &&\times \left. \int_0^1 \tilde{\Lambda}_i(u)\tilde{\Lambda}_j(u)^2 du \right] = 
3 \sum_{i = 0}^q \left[\int_0^1 \tilde{\Lambda}_i(u)du \right. \\&& \times \left. \int_0^1 \tilde{\Lambda}_i(u)u du\right]= \frac{1}{2}+\frac{1}{8} + 3 \sum_{i = 2}^q \left[\int_0^1 \tilde{\Lambda}_i(u)du \right. \\&& \times \left. \int_0^1 \tilde{\Lambda}_i(u)u du\right] = \frac{1}{2}+\frac{1}{8} ,
\end{eqnarray*}
where we used the equality
\[
\sum_{j = 0}^q \tilde{\Lambda}_j(u)^2 = u. 
\]
The above equation remains true of the discrete versions, too. For the full Brownian motion, one computes via the random series representation based on the Legendre orthogonal polynomials on the interval $[0,1]$ 
\begin{eqnarray*}&&
\mathbb{E}\left(\int_0^1 {B}_u du \int_0^1 {B}_u^3 du\right)= \frac{1}{2} + \frac{1}{8}
\end{eqnarray*}
and the fact that the equation $j_5 = 1, j_3 = 1$ is satisfied follows. 

We now turn our attention to the equation specified by $j_4 = 2$. One computes
\begin{eqnarray*}
\mathbb{E}\left(\int_0^1 \tilde{B}_u^2 du \right)^2 = \mathbb{E}\left[\int_0^1\left(\sum_{l = 0}^q a_i \tilde{\Lambda}_i(u)\right)^2du\right]^2 \nonumber \\ = 
\mathbb{E}\left(\sum_{i,j= 0 }^q a_i a_j c_{i,j}\right)^2, 
\end{eqnarray*}
where
\[c_{i,j} = \int_0^1 \tilde{\Lambda}_i(u)\tilde{\Lambda}_j(u) du.\]
Using Eq.~(\ref{eq:17}), one deduces
\begin{eqnarray*}
\mathbb{E}\left(\int_0^1 \tilde{B}_u^2 du \right)^2 = 2 \sum_{i,j=0}^q c_{i,j}^2 + \left(\sum_{i=0}^q c_{i,i}\right)^2.
\end{eqnarray*}

At this moment it is useful to remember that $\tilde{\Lambda}_0(u) = u$ and $\tilde{\Lambda}_1(u) = \sqrt{3}u(1-u)$. Moreover, notice that Eq.~(\ref{eq:II10a}) implies 
\[
\sum_{i = 0}^q c_{i,i} = \int_0^1[u^2 + u(1-u)]du = \frac{1}{2}.
\]
Therefore,
\begin{eqnarray*}&&
\mathbb{E}\left[\int_0^1 \tilde{B}_u^2 du \right]^2  = 2 \sum_{i,j=0}^q \left[ \int_0^1 \tilde{\Lambda}_i(u)\tilde{\Lambda}_j(u) du \right]^2 + \frac{1}{4}. \nonumber
\end{eqnarray*}

For the full Brownian motion, one computes via the Wiener-Fourier series
\begin{eqnarray*}
\mathbb{E}\left(\int_0^1 {B}_u^2 du \right)^2 =\frac{2}{9}+  4 \sum_{k=1}^\infty \left[\int_0^1 u\sqrt{\frac{2}{\pi^2}}\frac{\sin(k\pi u)}{k} du \right]^2 \\ + 2 \sum_{k=1}^\infty \left[ \int_0^1 \frac{2}{\pi^2}\frac{\sin(k\pi u)^2}{k^2} du \right]^2 + \frac{1}{4} = \frac{2}{9}+  \frac{8}{\pi^4} \sum_{k=1}^\infty \frac{1}{k^4}  \\ + \frac{2}{\pi^4} \sum_{k=1}^\infty \frac{1}{k^4} + \frac{1}{4} = \frac{2}{9} + \frac{1}{9} + \frac{1}{4}.
\end{eqnarray*}
Then, the equality 
\[\mathbb{E}\left(\int_0^1 \tilde{B}_u^2 du \right)^2 = \mathbb{E}\left(\int_0^1 {B}_u^2 du \right)^2\]
implies
\begin{equation}
\label{eq:18} 
\sum_{i,j=0}^q \left[ \int_0^1 \tilde{\Lambda}_i(u)\tilde{\Lambda}_j(u) du \right]^2 = \frac{1}{6}. 
\end{equation}
With the one-dimensional integrals replaced by appropriate quadrature sums, Eq.~(\ref{eq:18}) must also be satisfied by  all discrete short-time approximations of order $4$. Remember that the quadrature scheme is assumed to integrate exactly all the polynomials of degree at most $3$ and all the functions $\tilde{\Lambda}_k(u)$ for $2 \leq k \leq q$.

In the remainder of this subsection, we construct an example of reweighted short-time approximation of order $4$. Clearly, we cannot set $q = 2$ in Eq.~(\ref{eq:16}) because then 
\[
\tilde{\Lambda}_2(u)= \left\{u(1-u)[1-3u(1-u)]\right\}^{1/2},
\]
as follows from Eq.~(\ref{eq:II10a}), and consequently,
\[
\int_0^1 \tilde{\Lambda}_2(u) du \neq 0. 
\]
Thus, we set $q = 3$ and look for functions of the form
\begin{eqnarray}
\label{eq:19}
\left\{\begin{array}{c} 
\tilde{\Lambda}_2(u) = r(u) \cos[\alpha_1(u-0.5)+ \alpha_2(u-0.5)^3], \\ 
 \tilde{\Lambda}_3(u) = r(u) \sin[\alpha_1(u-0.5)+ \alpha_2(u-0.5)^3], 
\end{array}\right.
\end{eqnarray}
where
\[
r(u)= \left\{u(1-u)[1-3u(1-u)]\right\}^{1/2}.
\]
The functions $\tilde{\Lambda}_2(u)$ and $\tilde{\Lambda}_3(u)$ are orthogonal because the first is symmetric under the transformation $u'= 1- u$, whereas the second is antisymmetric. The integral over $[0,1]$ of the function $\tilde{\Lambda}_3(u)$ is zero by antisymmetry. 
Then, the constants $\alpha_1$ and $\alpha_2$ are determined from the system of equations
\begin{eqnarray}
\label{eq:20} \nonumber
1)&&  \int_0^1 \tilde{\Lambda}_2(u) du = 0,\\
2)&& \sum_{i,j=0}^3 \left[ \int_0^1 \tilde{\Lambda}_i(u)\tilde{\Lambda}_j(u) du \right]^2 = \frac{1}{6}.
\end{eqnarray}
The values of the constants $\alpha_1$ and $\alpha_2$ have been determined numerically to be
\begin{equation}
\label{eq:21}
\alpha_1 \approx 5.768064999 \quad \text{and} \quad \alpha_2 \approx 13.49214669.
\end{equation}

Let us now design a minimalist discrete short-time approximation of order $4$. Given an arbitrary symmetric quadrature technique that integrates exactly all polynomials of degree less or equal to $3$, we determine new values for $\alpha_1$ and $\alpha_2$ from the system of equations
\begin{eqnarray}
\label{eq:22} \nonumber
1)&&  \sum_{l= 1}^{n_q}w_l \tilde{\Lambda}_2(u_l) = 0,\\
2)&&  \sum_{i,j=0}^3 \left[ \sum_{l= 1}^{n_q}w_l\tilde{\Lambda}_i(u_l)\tilde{\Lambda}_j(u_l) \right]^2 = \frac{1}{6}.
\end{eqnarray}
Because there are two equations, it is easy to argue that the number of quadrature points lying in the open interval $(0,0.5)$ must be at least two. Consistent with this observation, the quadrature technique is chosen to be the $4$-point Gauss-Legendre technique on the interval $[0,1]$. This quadrature technique integrates exactly all the polynomials of degree at most $7$. The new values for the parameters $\alpha_1$ and $\alpha_2$ are then determined by solving the system of equations given by Eq.~(\ref{eq:22}) for the chosen quadrature scheme. The solution of the system of equations is given by
\begin{equation}
\label{eq:22a}
\alpha_1 \approx 6.379716466 \quad \text{and} \quad \alpha_2 \approx 8.160188248.
\end{equation}
The quadrature weights and points are presented in Table~\ref{Tab:III}, for ease of reference.
\begingroup
\begin{table}[!htbp]
\caption{
\label{Tab:III}
Quadrature points and weights for the minimalist discrete short-time approximation of order $4$. The quadrature points and weights are those for the $4$-point Gauss-Legendre technique on the interval $[0,1]$. }
\begin{tabular}{|c |c |c |c |c |}
\hline
$i$ & 1 & 2 & 3 & 4 \\ 
\hline \hline
$u_i$&0.069431844&0.330009478&0.669990522&0.930568156\\
\hline
$w_i$&0.173927423&0.326072577&0.326072577&0.173927423\\
\hline 
\end{tabular}
\end{table}
\endgroup

As shown by Eq.~(\ref{eq:II17}), the number of path variables entering the expression of $\rho_n(x,x';\beta)$ is $(q+1)n + q = 4n+3$, whereas the number of quadrature points [see Eq.~(\ref{eq:II18})] is $n_q(n+1)= 4n+4$. Thus, for large enough $n$, the ratio $(4n+4) / (4n+3)$ approaches $1$, value that equals the one for the trapezoidal Trotter discrete path integral method. Therefore, the fourth order method has the same numerical requirements as the trapezoidal Trotter discrete path integral method for equal numbers of path variables, yet it achieves quartic convergence for smooth enough potentials.

\section{Numerical verification of the asymptotic orders of convergence}

One of the main advantages of the Lie-Trotter product formula consists of the fact that, for low dimensional systems, the evaluation of the density matrix and related properties can be performed accurately by means of the numerical matrix multiplication (NMM) method.\cite{Kle73, Thi83} We shall use the NMM method to compute  $n$-th order approximations to the partition function of the type
\[
Z_n^{(\nu)}(\beta) = \int_{\mathbb{R}}\rho_n^{(\nu)}(x,x;\beta) d x, 
\]
for one-dimensional systems. We follow closely the simulation strategy employed in Ref.~\onlinecite{Pre03b} for a similar numerical study of asymptotic orders of convergence. The symbol $(\nu)$ to the exponent serves to differentiate between short-time approximations of different orders $\nu$. 

The main steps of the NMM algorithm are as follows. First, one restricts the system to an interval $[a, b]$ and considers a division of the interval of the type 
\[
x_i = a + i(b-a)/M, \quad 0 \leq i \leq M. 
\]
Next, one computes and stores the symmetric square  matrix of entries
\[A_{i,j} = \frac{b-a}{M} \rho_0^{(\nu)}\left(x_i, x_j ;\frac{\beta}{n+1}\right), \quad 0\leq i, j \leq M.\] The value of the partition function can then be recovered as 
\[
Z_n^{(\nu)}(\beta) = \text{tr}\left(A^{n+1}\right).
\]
By computer experimentation, the interval $[a, b]$ and the size $M$ of the division are chosen such that the computation of the partition function is performed with the required accuracy. A fast computation of the powers of the matrix $A$ can be achieved by exploiting the rule $A^{m+n}=(A^m)^n$.  For more details, the reader is referred to the cited literature. 

The Gaussian integrals appearing in the expression of the discrete reweighted short-time approximation
\begin{eqnarray*}
 \nonumber
\rho_0^{(\nu)}(x,x';\beta) = \rho_{fp}(x,x';\beta) \int_{\mathbb{R}} d \mu(a_1) \cdots \int_{\mathbb{R}} d \mu(a_q) \\ \times  \exp\left\{-\beta \sum_{i = 1}^{n_q}w_i V\left[x_r(u_i) + \sigma \sum_{k = 1}^q a_k \tilde{\Lambda}_k(u_i)\right]\right\}
\end{eqnarray*}
 can be evaluated by means of the Gauss-Hermite quadrature technique\cite{Pre92} for small enough $q$ (in our case, $q$ is $2$ for the approximation of order $3$ and $3$ for the approximation of order $4$, respectively).  For the purpose of establishing the asymptotic convergence of the partition functions, it was found that a number of $10$ quadrature points for each dimension is sufficient for both short-time approximations studied in the present section.  This is so because the errors due to the Gauss-Hermite quadrature approximation quickly vanish as $\beta / (n + 1) \to 0$. 
 
Once the partition functions are evaluated,  we compute the quantities
\begin{equation}
\label{eq:23}
R_{2m+1}^{(\nu)}(\beta) = Z_{2m+1}^{(\nu)}(\beta) \big/ Z(\beta)
\end{equation}
and
\[
\alpha_m^{(\nu)} = m^2 \ln\left[1+\frac{R_{2m-1}^{(\nu)}(\beta)-R_{2m+1}^{(\nu)}(\beta)} {R_{2m+1}^{(\nu)}(\beta)- 1}\right].
\]
As demonstrated in Ref.~\onlinecite{Pre02}, the slope of $\alpha_{m}^{(\nu)}$ as a function of $m$ converges to the convergence order. We want to verify whether or not this convergence order is $\nu$. The exact partition function $Z(\beta)$ necessary in Eq.~(\ref{eq:23}) is evaluated either by variational methods or by  employing a large $m$.

The first example studied is  the quartic potential $V(x)=x^4/2$. The following values of the physical constants (in atomic units) have been utilized: $\hbar =1$, $m_0=1$, and $\beta = 10$. The second example studied consists of a particle trapped on a line between two atoms separated by a distance $L$.\cite{Fre86} The particle is assumed to interact with the fixed atoms through pairwise Lennard-Jones potentials. The resulting cage is described by the potential
\[
V(x)= 4\epsilon \left[\left(\frac{\sigma}{x}\right)^{12} - \left(\frac{\sigma}{x}\right)^{6}+ \left(\frac{\sigma}{x-L}\right)^{12} - \left(\frac{\sigma}{x-L}\right)^{6}\right],
\]
if $0 < x < L$, and $V(x) = +\infty$, otherwise. The parameters of the system are chosen to be those for the He atom. We set $m_0 = 4\; \text{amu}$, $\epsilon / k_{B} = 10.22 \; \text{K}$, $\sigma = 2.556 \stackrel{\circ}{\text{A}}$, and $L= 7.153 \stackrel{\circ}{\text{A}}$. At $T = 5.11 \;\text{K}$, which is the temperature utilized in the present computations, the system is practically in its ground state. For more details regarding the present simulations, the reader is advised to consult Ref.~\onlinecite{Pre03b}.

\begin{figure}[!tbp] 
   \includegraphics[angle=270,width=9.5cm,clip=t]{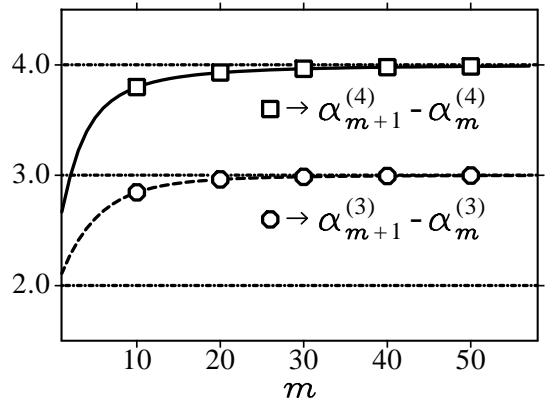} 
%    \BoxedEPSF{Potential.eps scaled 300}
 \caption[sqr]
{\label{Fig:1}
The convergence orders of the two discrete short-time approximations for the quartic potential. The plotting symbols are shown only for every tenth data point actually computed.
}
\end{figure} 

\begin{figure}[!tbp] 
   \includegraphics[angle=270,width=9.5cm,clip=t]{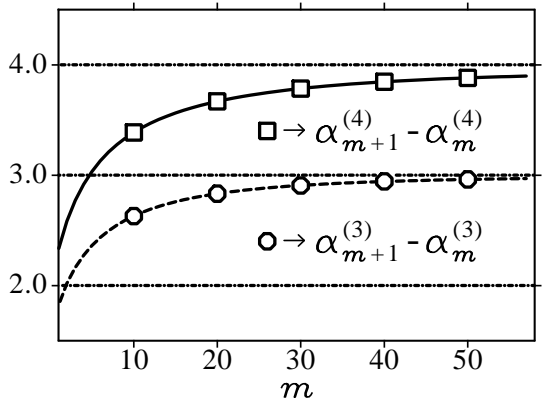} 
%    \BoxedEPSF{Potential.eps scaled 300}
 \caption[sqr]
{\label{Fig:2}
As in Fig.~\ref{Fig:1} for the He cage problem. 
}
\end{figure}

As Figs.~\ref{Fig:1} and~\ref{Fig:2} show, the orders of convergence predicted in the preceding section are well verified. I interpret these results as proof that the mathematical analysis performed in the present paper is sound. The He cage problem is interesting because the Lennard-Jones potential lies outside the class of potentials for which the theory was developed. As explained in Ref.~\onlinecite{Pre03b}, the  density matrix of the Lennard-Jones potential  has an exponential decay near singularities and therefore, the behavior of the potential near singularities is not important as far as the polynomial convergence of imaginary-time path integral methods is concerned. 

\section{Conclusions}

In this article, I have considered the problem of constructing direct short-time approximations to the density matrix of a physical system of arbitrary convergence orders. I have shown that the problem can be reduced to the construction of finite-dimensional approximations to the Brownian motion that  satisfy a certain system of functional equations. Using the developed theory, I have constructed two examples of reweighted short-time approximations having convergence orders $3$ and $4$, respectively. The predicted orders of convergence have been verified by numerical simulations. In Appendix~B, I have derived the convergence constant for the trapezoidal Trotter path integral method. The predicted convergence constant has also been verified by numerical simulations.

For imaginary-time path integral simulations, the reader may object that the use of a path integral technique having faster asymptotic convergence is not a significant algorithmic improvement because the final computational effort is eventually controlled by the rate of convergence of the Monte Carlo integration method. The computational effort, as measured against the number of calls to the potential $V(x)$, can be evaluated as follows. To attain a given absolute error $\epsilon$, one must utilize a number of 
\[
n = const / \epsilon^{1/\nu}
\]
path variables (here, $const$ is some proportionality constant). The cost to evaluate the average potential for a given path is equal to the number of quadrature points, which, in turn, is proportional to the number of path variables [here, we do not take into account the cost for the computation of the paths, which scales as $n \log_2(n)$, but which is usually negligible for the values of $n$ commonly employed in practice]. Thus, the cost for a single path evaluation is $const / \epsilon^{1/\nu}$. This cost is to be multiplied by the number of Monte Carlo steps, which is given by the formula
\[
No_{MC} = const / \epsilon^2,
\]
assuming that the variance of the Monte Carlo method does \emph{not} depend upon the number of path variables.
Thus, the total cost, defined as  the number of calls to the potential necessary to attain a given error, is 
\begin{equation}
\label{eq:45}
Cost = const \cdot  {\epsilon^{-(2 + 1/\nu)}},
\end{equation}
where $\nu$ is the convergence order of the \emph{direct} path integral method. Eq.~(\ref{eq:45}) shows that we cannot beat the slow convergence of the Monte Carlo integration scheme by increasing the order of convergence of the path integral technique. The total cost changes from $\epsilon^{-2.5}$ to $\epsilon^{-2.25}$ only, as we switch from the trapezoidal Trotter to the fourth order method designed in the present article. 

However, the methods designed in the present paper are still useful because the improvement, even if marginal, comes ``free of any charge.'' Indeed, as shown in Section~IV.B, the ratio number of quadrature points over number of path variables is $1$ (for $n$ large enough) for both the trapezoidal Trotter and the discrete fourth order method introduced in the present article. Therefore, there is no loss of efficiency in employing the discrete fourth order method even for those potentials for which the optimal convergence order is not attained. Because no additional cost is incurred even in the most disadvantageous situations, the discrete fourth order short-time approximation is a natural replacement for the trapezoidal Trotter short-time approximation in all path integral simulations.

At a more general level, the present development may be relevant for the problem of performing real-time path integral simulations.\cite{Dol99r} In this case, the asymptotic rate of convergence is \emph{crucial} because the noise in the Monte Carlo simulation not only that does depend upon the number of path variables, but actually increases exponentially fast with the number of path variables. This is the statement of the well-known dynamical sign problem.\cite{Ami93}

 Let us assume that for a given convergence order $\nu$, there is a finite system of functions $\{\tilde{\Lambda}_k(u); 0 \leq k \leq q_\nu\}$ that generates the short-time approximation of order $\nu$
\begin{eqnarray}
\label{eq:VI1} \nonumber
\rho_0^{(\nu)}(x,x';\beta) = \rho_{fp}(x,x';\beta) \int_{\mathbb{R}} d \mu_\sigma(a_1) \cdots \int_{\mathbb{R}} d \mu_\sigma(a_q) \\ \times  \exp\left\{-\beta \int_0^1 V\left[x_r(u) + \sum_{k = 1}^{q_\nu} a_k \tilde{\Lambda}_k(u)\right]du\right\}, \qquad
\end{eqnarray}
where 
\[
d\mu_\sigma (a_k)= (2\pi\sigma^2)^{-1/2} \exp\left[ - a_k^2/(2\sigma^2) \right] d a_k.
\]
Notice that in Eq.~(\ref{eq:VI1}) we have performed a substitution of variables $a'_k = \sigma a_k$ so that the dependence of the spread of the paths with $\beta$ is no longer buried in the potential [remember, $\sigma = (\hbar^2 \beta/m_0)^{1/2}$]. In principle, this transformation should allow us to extend the above formulas to complex-valued $\beta$. We ask the question of whether or not it is more optimal to give up the use of the Lie-Trotter product formula altogether and instead consider the sequence of approximations
\begin{equation}
\label{eq:VI2}
\rho_0^{(\nu)}(x,x';\beta) \to \rho(x,x';\beta)\quad \text{as} \quad \nu \to \infty.
\end{equation}
If with appropriate restrictions on $V(x)$ and $\psi(x)$ the series appearing in Eq.~(\ref{eq:5}) is analytic in $\beta$, it is straightforward to see that
\begin{equation}
\label{eq:VI3}
\int_\mathbb{R} \rho_0^{(\nu)}(x,x';\beta)\psi(x')dx' \to \int_\mathbb{R} \rho(x,x';\beta)\psi(x')dx' 
\end{equation}
exponentially fast as measured against $\nu$. 

It is then apparent that a favorable scaling of $q_\nu$ with $\nu$, as for instance a polynomial scaling, may strongly alleviate the dynamical sign problem. As the Hardy-Ramanujan formula shows, the number of equations that must by satisfied by the system of functions $\{\tilde{\Lambda}_k(u); 1\leq k \leq q_\nu\}$ increases with $\nu$ faster than any polynomial. However, this does not necessarily imply that $q_\nu$ increases with $\nu$ at the same rate. In the examples constructed in Section~IV, we have been able to accommodate the $18$ equations for order $3$ with only two functions, whereas the $40$ equations for order $4$ were accommodated with three functions. In both cases, the actual number of functions was much lower than the number of equations. I hope this short analysis justifies my belief that future research on the subject is worth the time of investigation and may lead to significant progress in the area of real-time path integral simulations.

\begin{acknowledgments} The author acknowledges support from the National
Science Foundation through awards Nos. CHE-0095053 and CHE-0131114. He
also wishes to thank Professor Jimmie D. Doll for helpful
discussions concerning the present work. Finally, the author would
like to express a special thanks to Dragos N. Oprea  for
pointing out the Hardy-Ramanujan asymptotic formula.
\end{acknowledgments}

\appendix
\section{Some mathematical facts about the Brownian motion and the Feynman-Kac formula}
In this appendix, I review the definition and some of the basic properties of the Brownian motion. In addition, an alternative formulation of the Feynman-Kac formula and the random series construction of the Brownian motion are presented. For further information, the reader is advised to consult the cited mathematical literature.\cite{Sim79, Dur96, McK69} Chapters I and II of Ref.~\onlinecite{Pre03g} also contain an in-depth introduction to Brownian motion and its relation to the Feynman-Kac formula.

A standard Brownian motion is defined as a stochastic process $\{B_u, u \geq 0\}$ that satisfies the following conditions:
\begin{enumerate}
\item[(a)]{Given $0\leq u_0 < u_1 < \ldots < u_n$ an arbitrary finite sequence of increasing times, the initial position $B_{u_0}$ and the position increments $B_{u_1} - B_{u_0},  \ldots, B_{u_n} - B_{u_{n-1}}$ are independent.}
\item[(b)]{If $s, u \geq 0$ and $[a,b]\subset \mathbb{R}$ is some arbitrary interval, then
\[P\left(B_{u+s} - B_u \in [a,b]\right) = \int_{a}^b \frac{1}{\sqrt{2\pi s}}\exp\left(-\frac{x^2}{2s}\right)dx.\]}
\item[(c)]{With probability one, the Brownian motion sampling paths $B_u$ are continuous.}
\end{enumerate}
The existence of a stochastic process satisfying the above conditions has been first proved by Wiener\cite{Wie23} in 1923.

If $B_0 = 0$ with probability one, then the Brownian motion is said to start at zero. In the present work, $B_u$ always denotes a standard Brownian motion starting at zero. The conditions (a) and (b) above are sufficient to demonstrate that the Brownian motion starting at zero is a Gaussian process with joint finite distributions given by \begin{eqnarray}
\label{eq:1.25} \nonumber &&
P\left(B_{u_1}\in [a_1, b_1], \ldots, B_{u_n}\in [a_n, b_n]\right) 
\\ && = \int_{a_1}^{b_1}dx_1 \cdots \int_{a_n}^{b_n} dx_n \prod_{i=1}^n p_{u_i - u_{i-1}}\left(x_{i-1}, x_i\right), \qquad
\end{eqnarray}
where $u_0 = 0$, $x_0 = 0$, and
\[
p_u(a,b) = \frac{1}{\sqrt{2\pi u}}\exp\left[-\frac{(b-a)^2}{2u}\right].
\]

Eq.~(\ref{eq:1.25}) can be utilized to compute the expected values of moments of standard Brownian motions starting at zero. For example,
\[
\mathbb{E}\left[(B_u)^4\right] = \int_{\mathbb{R}} \frac{1}{\sqrt{2\pi u}} \exp\left(-\frac{x^2}{2u}\right) x^4dx = 3u^2, 
\]
where we have used the fact that $B_u$ is a Gaussian variable centered about origin and of variance $u$, as follows from Eq.~(\ref{eq:1.25}). Therefore, 
\begin{equation}
\label{eq:B2}
\mathbb{E}\left[\int_0^1(B_u)^4du\right] = \int_0^1\mathbb{E}\left[(B_u)^4\right]du = \int_0^1 3u^2 du = 1. 
\end{equation}

A standard Brownian bridge $\{B_u^0, 0 \leq u \leq 1\}$ is defined as a standard Brownian motion starting at zero that is also conditioned to end up at zero at time $u = 1$:
\[
\{B_u^0, 0 \leq u \leq 1\} = \{B_u, 0 \leq u \leq 1 | B_1 = 0\}.
\]
A standard Brownian bridge can be  constructed from a standard Brownian motion starting at zero as the difference $B_u - u B_1$. More precisely, it can be demonstrated that
\[
\{B_u^0, 0 \leq u \leq 1 \}\stackrel{d}{=}\{B_u - u B_1, 0 \leq u \leq 1 \},
\]
where the symbol $\stackrel{d}{=}$ means that the left- and right-hand side processes are equal in distribution (have equal finite dimensional distributions) and have continuous sampling paths with probability one. Moreover, the random variables $B_1$ and $B_u^0 = B_u - u B_1$ are independent. It follows that given a Brownian bridge $B_u^0$ and an independent standard normal variable $z$ (which plays the role of $B_1$),  the sum of independent variables $B_u = B_u^0 + u z$ is equal in distribution to  a standard Brownian motion starting at zero. Thus,
\[
\left\{ B_u, 0 \leq u \leq 1\right\} \stackrel{d}{=} \left\{ B_u^0 + u z, 0 \leq 1 \leq u\right\}
\] 
and $z \stackrel{d}{=} B_1$ (because $B_1^0 = 0$, by the very definition of the Brownian bridge). 

As Simon often emphasizes,\cite{Sim79} Eq.~(\ref{eq:I1}) presented in the introduction is only one of the many equivalent formulations of the Feynman-Kac formula. Another popular formulation, which utilizes the full Brownian motion rather than the Brownian bridge, will be presented shortly. Let $\psi(x)$ be an arbitrary square integrable function. From Eq.~(\ref{eq:I1}), we have
\begin{eqnarray*}&&
\left\langle x \left| e^{-\beta H} \right| \psi \right \rangle = \int_{\mathbb{R}} dx'  \frac{1}{\sqrt{2\pi\sigma^2}} \exp\left[-\frac{(x'-x)^2}{2\sigma^2}\right]  \\ && \times \mathbb{E}\exp\left\{-\beta\int_{0}^{1}\! \!  V\Big[x+(x'-x)u+\sigma B_u^0 \Big] d u\right\} \psi(x').
\end{eqnarray*}
Performing the substitution $x' = x + \sigma z$, we obtain
\begin{eqnarray*}&&
\left\langle x \left| e^{-\beta H} \right| \psi \right \rangle =  \int_{\mathbb{R}} dz \mathbb{E}  \frac{1}{\sqrt{2\pi}} \exp\left(-{z^2}/{2}\right)  \\ && \times \exp\left\{-\beta\int_{0}^{1}\! \!  V\Big[x+\sigma z u+\sigma B_u^0 \Big] d u\right\} \psi(x + \sigma z). 
\end{eqnarray*}
Notice that the variables $z$ and $B_u^0$, as they appear in the preceding equation, are independent. Moreover,  $z$ is a Gaussian variable centered in zero and of mean $1$. It follows that $zu + B_u^0$ is equal in distribution to a Brownian motion $B_u$ starting at zero. In these conditions,  the Feynman-Kac formula reads
\begin{equation}
\label{eq:B4}
\left\langle x \left| e^{-\beta H} \right| \psi \right \rangle =   \mathbb{E} \left[  e^{-\beta\int_{0}^{1}\! \!  V\left(x+\sigma B_u \right) d u} \psi(x + \sigma B_1) \right],
\end{equation}
where the symbol $\mathbb{E}$ denotes the expected value with respect to the entire Brownian motion $B_u$.

I conclude this appendix by presenting the statement of the Ito-Nisio theorem,\cite{Kwa92, Pre03g} theorem that gives an explicit construction of a standard Brownian motion over the interval $[0,1]$ as a random series. 
\begin{5}[Ito-Nisio]
\label{th:5}
Let $\{\lambda_k(\tau)\}_{k \geq 0}$ be any orthonormal basis in $L^2[0,1]$, let
\[
\Lambda_k(u)=\int_{0}^{u}\lambda_k(\tau)d\tau, 
\]
and let $\bar{a} := \{a_0, a_1, \ldots\}$ be a sequence of 
 distributed  standard normal random variables. Then, the random series $\sum_{k=0}^\infty a_k \Lambda_k(u)$ is uniformly convergent almost surely  and equal in distribution  over the interval $[0,1]$ with a standard Brownian motion $B_u$ starting at zero.
\end{5}

To express the Feynman-Kac formula as the expected value of a functional of a random series, it is convenient to work with those
  orthonormal basis $\{\lambda_k(\tau)\}_{k \geq 0}$ for which $\lambda_0(\tau) = 1$ only. Then $\Lambda_0(u) = u$ and 
\[
\Lambda_k(1) = \int_0^1 \lambda_k(\tau)d\tau = \int_0^1 \lambda_k(\tau)\lambda_0(\tau)d\tau = 0
\]
for all $k \geq 1$. In these conditions, the Ito-Nisio theorem says that 
\[
 \sum_{k = 1}^\infty a_k \Lambda_k(u) =  \sum_{k = 0}^\infty a_k \Lambda_k(u) - a_0 u \stackrel{d}{=} B_u - u B_1.
\]
The last term in the preceding equation has been discussed in a previous paragraph to be equal in distribution to a Brownian bridge. It follows that if $\lambda_0(\tau) = 1$, then
\[
B_u^0 \stackrel{d}{=} \sum_{k = 1}^\infty a_k \Lambda_k(u), \quad 0 \leq u \leq 1,
\]
equality in distribution that provides an explicit random series construction for the standard Brownian bridge. 

In these conditions, if $\Omega$ is the set of all sequences $\bar{a} := \{a_1, a_2, \ldots\}$ and
 if 
\[
dP[\bar{a}] = \prod_{k = 1}^\infty \frac{1}{\sqrt{2\pi}}e^{-a_k^2/2}da_k
\]
is the probability measure on $\Omega$ associated with the sequence of independent random variables $\bar{a} := \{a_1, a_2, \ldots\}$, then the Feynman-Kac formula given by Eq.~(\ref{eq:I1}) reads
\begin{eqnarray}
\label{eq:B5}\nonumber&&
\frac{\rho(x,x';\beta)}{\rho_{fp}(x,x';\beta)}=\int_{\Omega}dP[\bar{a}] \\ && \times  \exp\left\{-\beta\int_{0}^{1}\! \!  V\left[x_r(u)+\sigma \sum_{k = 1}^\infty a_k \Lambda_k(u)\right] d u\right\}. \qquad
\end{eqnarray}
Eq.~(\ref{eq:B5}) is called the random series representation of the Feynman-Kac formula.\cite{Pre02} 

\section{The convergence constant for the trapezoidal Trotter approximation}
The short-time approximation for the trapezoidal Trotter path integral method is given by the expression
\[
\rho_0^{\text{TT}}(x,x';\beta) = \rho_{fp}(x,x';\beta) \exp\left[-\beta\frac{V(x)+V(x')}{2}\right].
\]
This short-time approximation is of the type given by Eq.~(\ref{eq:II16}), provided that the quadrature technique is specified by the two points $u_0 = 0$ and $u_1 = 1$, and the weights $w_0 = 1/2$ and $w_1 = 1/2$, respectively. The approximation is independent of the functions $\{\tilde{\Lambda}_k(u); 0 \leq k \leq q\}$, because the end points of these functions are specified by Eq.~(\ref{eq:II10}). We can therefore consider that the functions are those for the third order reweighted approximation, or one may work with a full random series representation of the Brownian motion of the type
\[
a_0 u + \sum_{k = 1}^\infty a_k \Lambda_k(u),
\]
as provided by the Ito-Nisio theorem. It does not make any difference. The trapezoidal Trotter approximation is just a discrete version of the third order  reweighted technique discussed in Section~IV.A or of the full Feynman-Kac formula.

Using the fact that the trapezoidal quadrature rule given above integrates exactly the polynomials $1$ and $u$, the reader may argue that all equations specified in Table~\ref{Tab:I} with $\mu = 1, 2, 3$ are satisfied, except for the following (for all, $\mu = 3$):

1) Case $j_6 = 1$. For the full Brownian motion, one computes
\begin{eqnarray*}
\mathbb{E}(M_4) = \mathbb{E}\int_0^1 (B_u)^4du =\int_0^1 \mathbb{E}(B_u)^4du  =3\int_0^1 u^2du = 1.
\end{eqnarray*}
The trapezoidal rule produces the different result
\[
\frac{3}{2}(0^2 + 1^2) = \frac{3}{2}.
\]

2) Case $j_5 = 1$ and $j_1 = 1$. For the full Brownian motion, one computes
\[
\mathbb{E}(B_1M_3)=\mathbb{E}\left[B_1\int_0^1 (B_u)^3du\right] = 3\int_0^1u^2du = 1.
\]
The trapezoidal rule produces 
\[
3\frac{1}{2}(0^2 + 1^2) = \frac{3}{2}.
\]

3) Case $j_4 = 1$ and $j_1 = 2$. For the full Brownian motion, we have
\begin{eqnarray*}
\mathbb{E}[(B_1)^2M_2] = \mathbb{E}\left[(B_1)^2\int_0^1 (B_u)^2du\right] \\ = \int_0^1(2u^2+u)du = \frac{2}{3}+\frac{1}{2}.
\end{eqnarray*}
The trapezoidal rule produces
\[
\frac{1}{2}(2+1)= \frac{3}{2}.
\]

4) Case $j_3 = 2$. For the full Brownian motion, we have [see Eq.~(\ref{eq:11})]
\[
\mathbb{E}[(M_1)^2] = \mathbb{E} \left(\int_0^1 B_u du \right)^2 =\frac{1}{4}+ \frac{1}{12}.
\]
The trapezoidal rule produces
\[
\mathbb{E} \left[\frac{1}{2}(0 + B_1)\right]^2 = \frac{1}{4}.
\]

With the help of the series given by Eqs.~(\ref{eq:5}) and (\ref{eq:6}), we compute 
\begin{eqnarray*}&&
\int_\mathbb{R}\left[\rho^{\text{TT}}_0(x,x';\beta)- \rho(x,x';\beta)\right]\psi(x')dx' =
\beta^3 \\&& \times \bigg\{ \frac{(-1)^1}{4!}\left(\frac{3}{2}-1\right)\left(\frac{\hbar^2}{m_0}\right)^2V^{(4)}(x)\psi(x) \\&& + \frac{(-1)^1}{3!}\left(\frac{3}{2}-1\right)\left(\frac{\hbar^2}{m_0}\right)^2V^{(3)}(x)\psi^{(1)}(x) \\&& + 
\frac{(-1)^1}{2!2!}\left(1-\frac{2}{3}\right)\left(\frac{\hbar^2}{m_0}\right)^2V^{(2)}(x)\psi^{(2)}(x)\\&& +
\frac{(-1)^2}{2!}\left(-\frac{1}{12}\right)\frac{\hbar^2}{m_0}\left[V^{(1)}(x)\right]^2\psi(x)\bigg\} + O(\beta^4).
\end{eqnarray*}

From the equation above, we learn that the trapezoidal Trotter path integral technique has convergence order $2$. Moreover, the convergence operator for the trapezoidal Trotter short-time approximation is \begin{eqnarray}
\label{eq:A1}\nonumber
T_2 =  -\frac{1}{48}\left(\frac{\hbar^2}{m_0}\right)^2V^{(4)}(x) -
\frac{1}{24}\frac{\hbar^2}{m_0}\left[V^{(1)}(x)\right]^2 \\ -\frac{1}{12}\left(\frac{\hbar^2}{m_0}\right)^2\frac{d}{dx} \left(
V^{(2)}(x)\frac{d}{dx}\right).
\end{eqnarray}
The above form of Eq.~(\ref{eq:A1}) emphasizes the Hermiticity of the convergence operator. According to Theorem~\ref{th:2}, the following result is expected to hold. 
\begin{6}
\label{th:6}
The convergence constant for the trapezoidal Trotter path integral method is given by the formula
\begin{eqnarray}
\label{eq:A2}\nonumber
\lim_{n\to \infty} {(n+1)^2}\left[\rho^{\text{TT}}_n(x,x';\beta)- \rho(x,x';\beta)\right]= \\ {\beta^{3}} \int_{0}^1\left\langle x\left|e^{-\theta \beta H}T_2 e^{-(1-\theta)\beta H}\right|x'\right\rangle 
d \theta, 
\end{eqnarray}
where the operator $T_2$ is defined by Eq.~(\ref{eq:A1}).
\end{6}

For the purpose of numerical verification, we derive the convergence constant for the partition function. Though one can work with the full density matrix and employ the Bloch equation whenever necessary, it seems that it is more convenient to utilize an eigenfunction expansion for the density matrix.  Setting $x' = x$ and integrating over $x$ in Eq.~(\ref{eq:A2}), we obtain, after several simplifications and an integration by parts,
\begin{eqnarray}
\label{eq:A3} \nonumber
\lim_{n\to \infty} {(n+1)^2}\left[Z^{\text{TT}}_n(\beta)- Z(\beta)\right] = \frac{1}{24} \frac{\hbar^2\beta^{3}}{m_0} \sum_{k = 0}^\infty  e^{-\beta E_k}\\
 \times \int_\mathbb{R} \bigg\{-\frac{\hbar^2}{2m_0} V^{(4)}(x) \psi_k(x)^2 -  \left[V^{(1)}(x)\right]^2 \qquad \\
  \times \psi_k(x)^2   + \frac{2\hbar^2}{m_0} V^{(2)}(x) \left[\frac{d}{dx}\psi_k(x)\right]^2 \bigg\}dx.
\nonumber
\end{eqnarray}

Integrating by parts three times, one argues that 
\begin{eqnarray*}&&
-\frac{\hbar^2}{2m_0}\int_\mathbb{R} V^{(4)}(x) \psi_k(x)^2dx = \frac{\hbar^2}{m_0} \int_\mathbb{R} V^{(1)}(x) \\ &&
 \times \frac{d}{dx}\left\{ \left[\psi_k(x)\frac{d^2}{dx^2}\psi_k(x)\right] + \left[\frac{d}{dx}\psi_k(x)\right]^2\right\}dx,
\end{eqnarray*}
whereas, integrating by parts once, we obtain
\begin{eqnarray*}
\frac{2\hbar^2}{m_0}\int_\mathbb{R} V^{(2)}(x) \left[\frac{d}{dx}\psi_k(x)\right]^2 dx = - \frac{2\hbar^2}{m_0} \\ \times \int_\mathbb{R} V^{(1)}(x)  \frac{d}{dx}  \left[\frac{d}{dx}\psi_k(x)\right]^2dx.
\end{eqnarray*}
Adding the last two equations and simplifying, we get
\begin{eqnarray}
\label{eq:A4}&& \nonumber
-\frac{\hbar^2}{2m_0}\int_\mathbb{R} V^{(4)}(x) \psi_k(x)^2dx +   \frac{2\hbar^2}{m_0}\int_\mathbb{R} V^{(2)}(x) \\
&& \times \left[\frac{d}{dx}\psi_k(x)\right]^2 dx = \frac{\hbar^2}{m_0} \int_\mathbb{R} V^{(1)}(x) \\
&& \times \left\{ \psi_k(x)\frac{d^3}{dx^3}\psi_k(x) - \left[\frac{d}{dx}\psi_k(x)\right]\left[\frac{d^2}{dx^2}\psi_k(x)\right]\right\}dx, \nonumber
\end{eqnarray}

However, by virtue of the Schr{\"o}dinger equation, we have the equality
\begin{eqnarray*}
-\frac{\hbar^2}{2m_0} \left\{ \psi_k(x)\frac{d^3}{dx^3}\psi_k(x) - \left[\frac{d}{dx}\psi_k(x)\right]\left[\frac{d^2}{dx^2}\psi_k(x)\right]\right\} \\ =
\psi_k(x)\frac{d}{dx}\left\{ [E_k - V(x)]\psi_k(x)\right\} -  \left[\frac{d}{dx}\psi_k(x)\right] \\ \times \left\{[E_k - V(x)]\psi_k(x)\right\} =
- \psi_k(x)^2 V^{(1)}(x).
\end{eqnarray*}
Replacing the last equality in Eq.~(\ref{eq:A4}), we obtain
\begin{eqnarray*}
&& \nonumber
-\frac{\hbar^2}{2m_0}\int_\mathbb{R} V^{(4)}(x) \psi_k(x)^2dx +   \frac{2\hbar^2}{m_0}\int_\mathbb{R} V^{(2)}(x) \\
&& \times \left[\frac{d}{dx}\psi_k(x)\right]^2 dx = 2 \int_\mathbb{R} \left[V^{(1)}(x)\right]^2\psi_k(x)^2 dx, 
\end{eqnarray*}
relation that, upon substitution in Eq.~(\ref{eq:A3}), produces the following corollary of Theorem~\ref{th:5}.
\begin{Cor1}
\label{co:1}
The convergence constant for the relative error of the partition function for the trapezoidal Trotter path integral technique is given by the average
\begin{eqnarray}
\label{eq:A5}\nonumber
\lim_{n\to \infty} {(n+1)^2}\frac{Z^{\text{TT}}_n(\beta)- Z(\beta)}{Z(\beta)}= \\ \frac{1}{24} \frac{\hbar^2\beta^{3}}{m_0} \frac{\int_\mathbb{R} \left[V^{(1)}(x)\right]^2\rho(x;\beta) dx}{\int_\mathbb{R} \rho(x;\beta) dx},
\end{eqnarray}
where $\rho(x;\beta) = \rho(x,x;\beta)$ is the diagonal density matrix.
\end{Cor1}

\emph{Observation.} It can be shown that for a multidimensional system, the convergence constant is given by the formula
\begin{eqnarray}
\label{eq:A6}\nonumber &&
\lim_{n\to \infty} {(n+1)^2}\frac{Z^{\text{TT}}_n(\beta)- Z(\beta)}{Z(\beta)}= \\ && \frac{\hbar^2\beta^{3}}{24} \sum_{i = 1}^{d}\frac{1}{m_{0,i}} \frac{\int_{\mathbb{R}^d}  \left[\partial_i V(x)\right]^2\rho(x;\beta) dx}{\int_{\mathbb{R}^d} \rho(x;\beta) dx},
\end{eqnarray}
where $\partial_i V(x)$ denotes the partial derivative with respect to the coordinate $i$.

\begin{figure}[!tbp] 
   \includegraphics[angle=270,width=9.5cm,clip=t]{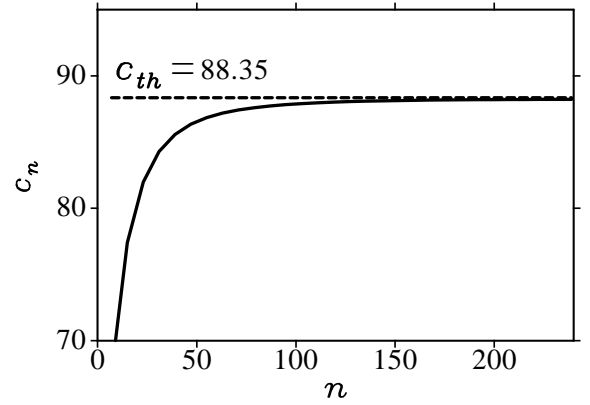} 
%    \BoxedEPSF{Potential.eps scaled 300}
 \caption[sqr]
{\label{Fig:3}
The convergence constant for the relative error of the partition function of the quartic oscillator, computed for the trapezoidal Trotter path integral method. The sequence of observed convergence constants $c_n$ is seen to converge to the theoretical value of $c_{th} \approx 88.35$, which is the value predicted by Corollary~\ref{co:1}.
}
\end{figure} 

\begin{figure}[!tbp] 
   \includegraphics[angle=270,width=9.5cm,clip=t]{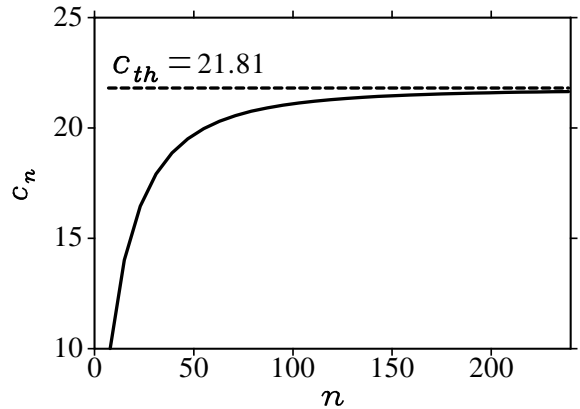} 
%    \BoxedEPSF{Potential.eps scaled 300}
 \caption[sqr]
{\label{Fig:4}
Same as in Fig.~\ref{Fig:3}, but for the He cage problem.
}
\end{figure} 

The numerical verification of Corollary~\ref{co:1} is done by numerical matrix multiplication for the systems discussed in Section~V.  The theoretical convergence constants
\[
c_{\text{th}} = \frac{1}{24} \frac{\hbar^2\beta^{3}}{m_0} \frac{\int_\mathbb{R} \left[V^{(1)}(x)\right]^2\rho(x;\beta) dx}{\int_\mathbb{R} \rho(x;\beta) dx}
\]
can also be computed by numerical matrix multiplication (or, more generally, by Monte Carlo integration). The experimental values are obtained by numerically studying the limit of the sequence
\[
c_n = {(n+1)^2}\frac{Z^{\text{TT}}_n(\beta)- Z(\beta)}{Z(\beta)}.
\] 
As Figs.~\ref{Fig:3} and \ref{Fig:4} show, the agreement between the theoretical and the experimentally observed convergence constants is excellent for both  the quartic oscillator and  the He cage problem.
This agreement is further evidence that the statement of Theorem~\ref{th:2} is correct, at least for the class of potentials and short-time approximations considered in the present article.

\end{document}